\documentclass[10pt,a4paper]{article}

\usepackage[T1]{fontenc}
\usepackage[utf8]{inputenc}
\usepackage{amssymb,amsmath,amsthm,amsfonts}    
\usepackage{bm}                                 
\usepackage{mathtools}                          
\usepackage{stmaryrd}                           
\usepackage{breqn}                              
\usepackage[textfont=footnotesize,labelfont={footnotesize,bf}]{caption}
\usepackage{subcaption}
\usepackage{booktabs}
\usepackage[inline]{enumitem}
\usepackage{graphicx}
\usepackage[dvipsnames]{xcolor}
\usepackage[top=30mm,right=30mm,left=30mm,bottom=30mm]{geometry}
\usepackage[colorlinks=true,linkcolor=blue,urlcolor=blue]{hyperref}
\usepackage[affil-it,auth-sc]{authblk}          
\usepackage[colorinlistoftodos,textwidth=25mm,textsize=footnotesize]{todonotes}
\usepackage{lipsum}                             
\usepackage[
    backend=biber,
    style=numeric,
    citestyle=numeric-comp,
    maxbibnames=99,
    minbibnames=99,
    maxcitenames=2,
    mincitenames=1,
    giveninits=true,
    sorting=none,
    sortlocale=auto,
    natbib=true,
    url=false,
    isbn=false,
    doi=true,
    eprint=true
]{biblatex}
\DeclareMathOperator{\curl}{curl}

\DeclareMathOperator{\Rp}{Re}
\DeclareMathOperator{\Ip}{Im}
\DeclarePairedDelimiter{\abs}{\lvert}{\rvert}
\DeclarePairedDelimiter{\norm}{\lVert}{\rVert}

\newcommand{\scalp}{\bm \cdot}
\newcommand{\trans}[1]{#1^{\mathsf{T}}}
\newcommand{\conj}[1]{\overline{#1}}
\newcommand{\conjtrans}[1]{#1^{\mathsf{H}}}
\newcommand{\deriv}[2]{\frac{\partial #1}{\partial #2}}

\newcommand{\Reals}{\mathbb{R}}
\newcommand{\Complexes}{\mathbb{C}}

\newcommand{\bzero}{\bm 0}
\newcommand{\ba}{\bm a}

\newcommand{\be}{\bm e}
\newcommand{\bef}{\bm f}
\newcommand{\bg}{\bm g}
\newcommand{\bh}{\bm h}

\newcommand{\bk}{\bm k}

\newcommand{\bn}{\bm n}

\newcommand{\bq}{\bm q}

\newcommand{\bt}{\bm t}
\newcommand{\bu}{\bm u}

\newcommand{\by}{\bm y}

\newcommand{\bA}{\bm A}

\newcommand{\bI}{\bm I}

\newcommand{\bK}{\bm K}
\newcommand{\bL}{\bm L}
\newcommand{\bM}{\bm M}
\newcommand{\bN}{\bm N}

\newcommand{\bP}{\bm P}
\newcommand{\bQ}{\bm Q}

\newcommand{\bS}{\bm S}

\newcommand{\bZ}{\bm Z}

\newcommand{\btau}{\bm \tau}

\newcommand{\bGamma}{\bm \Gamma}

\newcommand{\bXi}{\bm \Xi}
\newcommand{\bPi}{\bm \Pi}

\newcommand{\bUpsilon}{\bm \Upsilon}

\newcommand{\bPsi}{\bm \Psi}

\newcommand{\fH}{\mathbb H}

\newcommand{\mB}{\mathcal B}
\newcommand{\mC}{\mathcal C}

\newcommand{\mF}{\mathcal F}

\newcommand{\mL}{\mathcal L}

\newcommand{\mO}{\mathcal O}

\newcommand{\mT}{\mathcal T}

\newcommand{\mV}{\mathcal V}
\newcommand{\mW}{\mathcal W}

\graphicspath{{./}{./figures/}}                 

\title{A way to hypo-elastic artificial materials without a strain potential and displaying flutter instability}

\author[1]{G. Bordiga}
\author[1]{A. Piccolroaz}
\author[1]{D. Bigoni\footnote{Corresponding author: e-mail: \href{mailto:bigoni@ing.unitn.it}{bigoni@ing.unitn.it}; phone: +39\,0461\,282507.}}

\affil[1]{Department of Civil, Environmental, and Mechanical Engineering, University of Trento, Trento, Italy}

\date{\today} 

\begin{document}

\maketitle

\begin{abstract}
    \noindent
    Cauchy-elastic solids include hyper-elasticity and a subset of elastic materials for which the stress does not follow from a scalar strain potential.
    More in general, hypo-elastic materials are only defined incrementally and comprise Cauchy-elasticity.
    Infringement of the hyper-elastic `dogma' is so far unattempted and normally believed to be impossible, as it apparently violates thermodynamics, because energy may be produced in closed strain cycles.
    Contrary to this belief, we show that non-hyper-elastic behavior is possible and we indicate the way to a practical realization of this new concept.
    In particular, a design paradigm is established for artificial materials where follower forces, so far ignored in homogenization schemes, are introduced as loads prestressing an elastic two-dimensional grid made up of linear elastic rods (reacting to elongation, flexure and shear).
    A rigorous application of Floquet-Bloch wave asymptotics yields an unsymmetric acoustic tensor governing the \textit{incremental} dynamics of the effective material.
    The latter is therefore the incremental response of a hypo-elastic solid, which does \textit{not} follow from a strain potential and thus does \textit{not} belong to \textit{hyper-elasticity}.
    Through the externally applied follower forces (which could be originated via interaction with a fluid, or a gas, or by application of Coulomb friction, or non-holonomic constraints), the artificial material may adsorb/release energy from/to the environment, and therefore produce energy in a closed strain loop, without violating any rule of thermodynamics.
    The solid is also shown to display flutter, a material instability corresponding to a Hopf bifurcation, which was advocated as possible in plastic solids, but never experimentally found and so far believed to be impossible in elasticity.
    The discovery of elastic materials capable of sucking up or delivering energy in closed strain cycles through interaction with the environment paves the way to realizations involving micro and nano technologies and finds definite applications in the field of energy harvesting.
\end{abstract}

\paragraph{Keywords}
Cauchy-elasticity \textperiodcentered\
Follower forces \textperiodcentered\
Dynamic homogenization \textperiodcentered\
Hypo-elasticity \textperiodcentered\
Bloch waves

\section{Introduction}
\label{sec:introduction}
In a \textit{Cauchy-elastic} solid, the state of stress at a point in the current configuration remains determined by the state of strain at that point, measured with respect to an arbitrarily chosen reference configuration.
Consequently, the Cauchy stress does not depend on the deformation path followed from the reference configuration, even if the work done by the stress may be path-dependent.
When it is not, the material is called \textit{hyper-elastic} and the stress can be derived from a strain potential.
Therefore, Cauchy-elasticity defines a set of constitutive responses broader than hyper-elasticity, thus comprising materials for which a strain potential cannot be defined.

A constitutive response more general than Cauchy-elasticity is \textit{hypo-elasticity}, defined only incrementally and with the sole constraint that strain is \textit{incrementally recovered} upon unloading.
Therefore, the elastic fourth-order tensor lacks the major symmetry and a strain potential does not exist, except for special classes of material responses, for instance hyper-elasticity~\cite{truesdell_1955,truesdell_2004}.

In a linearized theory, the incremental stress is merely determined by the incremental strain and, when the transformation is non-Hermitian, the potential is absent, so that the material can produce or absorb energy when subject to a closed loop of strain.
The latter circumstance leads to possible dynamic material instabilities and is believed to preclude the existence of a \textit{passive} material of that nature, which would violate general restrictions from thermodynamics in an isolated system.
This belief is rooted in the beginning of the nineteenth century, when George Green~\cite{green_1839} established the symmetry of the elastic laws\footnote{
Lord Kelvin, (1883, Sect. 673)~\cite{thomson_1883} writes ``[The linear elastic stress-strain] equations express the six components of stress as linear functions of the six components of strain with 15 equalities among their 36 coefficients, which leave only 21 of them independent. [...]
But it is only by the principle of energy that, as first discovered by Green, the fifteen pairs of these coefficients are proved to be equal''.
}
and Lord Kelvin~\cite{thomson_1855} pointed out the concept of strain energy function for elastic materials\footnote{
    Lord Kelvin (1855) writes ``the work required to strain the body from one to another of two given mechanical states, keeping it always at the same temperature, is independent of the particular succession of mechanical states through which it is made to pass, and is always the same when the initial and final states are the same.
    This theorem was, I believe, first given by Green (as a consequence, I suppose, of the most general conceivable hypothesis that could be framed to explain the mutual actions of the different parts of a body on which its elasticity depends), who inferred from it that there cannot be 36, but only 21, independent coefficients of elasticity [...].
    It is now demonstrated as a particular consequence of the Second General Thermodynamic Law''.
}, called now hyper-elastic.

While hyper-elasticity has become a reference theory and elastic materials are ubiquitous, hypo-elasticity and Cauchy-elasticity not following from a potential have been abandoned, so that the latter materials are unknown and believed to be excluded from the real world.
Despite this belief, a rigorous application of homogenization theory is provided in the present article to show that
\begin{quote}
    \textit{passive elastic materials without a strain energy potential are possible}.
\end{quote}
In particular, Floquet-Bloch wave asymptotics is used to find the effective behavior for the in-plane \textit{incremental} response of a periodic grid of elastic rods, prestressed by \textit{follower} forces.
From these forces, generated as an the interaction with an `environment',  the material can absorb energy, so that production or release of work in a closed strain cycle does not violate thermodynamics.
The effective material is given through the explicit, closed-form calculation of its \textit{unsymmetric} acoustic tensor.
This unsymmetry proves that the effective elastic material does not admit a strain potential and thus is a hypo-elastic or a Cauchy-elastic material.

The effective hypo-elastic material is found to suffer instabilities, in particular, flutter instability, a Hopf bifurcation which was advocated~\cite{rice_1976}, although never indisputably experimentally found, to be possible in plastic materials, but \textit{not} in elastic
ones\footnote{
    Among the material instabilities that can be captured by continuum models, flutter instability still remains the most elusive.
    A continuum material capable of displaying such dynamic instability has been shown to be possible through a non-associative elastoplastic constitutive law~\cite{bigoni_1995,bigoni_1999,piccolroaz_2006}.
    However, the origin of the constitutive law used so far is phenomenological, so that connections between constitutive parameters and microstructure are not clearly determined. Moreover, flutter is predicted to occur considering only the loading branch of the piece-wise linear constitutive operator of plasticity, so that the real nonlinear behavior is unknown.
    The present study proves the existence of an alternative way to achieve flutter instability through an \textit{elastic} material.
}.
Remarkably, flutter instability is found to occur under a \textit{tensile} follower load, a result so far believed to be achievable only through the use of sliding constraints, in analogy with the buckling in tension~\cite{zaccaria_2011}.
The development of flutter instability is investigated by comparing the response of the periodic lattice to a pulsating concentrated force (analyzed numerically via the finite element method) with its corresponding effective hypo-elastic medium (analyzed analytically with the relevant infinite-body dynamic Green's function \cite{piccolroaz_2006}).
This comparison, never so far attempted, shows that the blowing-up nature of flutter can be replicated in an artificial elastic material, when subject to follower forces.

It is noted in closure that the results presented in this article share some common points with the so-called `non-Hermitian systems' investigated in the dynamics of metamaterials.
These materials can exchange energy with their surroundings, but have been introduced as  \textit{active}~\cite{torrent_2018,torrent_2018a,wang_2018,chen_2019,attarzadeh_2020,rosa_2020}, so that they contain motors, actuators, or piezoelectric devices, which can apply non-trivial forces or change the properties of the elements internal to the system itself.
Moreover, many of the realizations so far presented are unidimensional and no attempt has ever been made to derive an effective material behavior in which the elastic tensor lacks the major symmetry.
A more detailed presentation of results already available is postponed to Appendix~\ref{sec:non-hermitian_literature}, where the interested reader is referred.

The architected elastic materials explored in the present article are passive, but can absorb and release energy when subject to strain cycles of non-null area, an effect which is here related to the concept of follower force.
The passive nature of these materials lies in the fact that follower loads can be applied to an elastic structure through passive devices by exploiting friction or non-holonomic constraints~\cite{bigoni_2018a,cazzolli_2020}.
This is in stark contrast with previously mentioned results on non-Hermitian systems, which rely on a careful tuning of feedback controls, and thus are relevant for modeling active systems, but not for energy harvesting applications.
Therefore, the combination of the homogenization approach presented in this article with the loading strategies recently pointed out to generate follower forces~\cite{bigoni_2018a,cazzolli_2020} leads to the design of a new class of materials, never so far explored and waiting for applications in soft devices and metamaterials, capable of harnessing mechanical energy from the environment.

The article is organized as follows.
Section~\ref{sec:energy_loop} demonstrates the link between lack of major symmetry in the elasticity tensor and the capability of a material to absorb/release energy during closed deformation paths.
The formulation for the non-Hermitian incremental dynamics of lattices, preloaded by follower forces, is introduced in Section~\ref{sec:problem_setting}, while the homogenization technique leading to the hypo-elastic effective material is developed in Section~\ref{sec:homogenization}.
A concrete example showcasing an anisotropic lattice is presented in Section~\ref{sec:grid_example}, where the dynamics and stability of both the effective medium and the lattice are analyzed.
Section~\ref{sec:flutter_forced} investigates the manifestation of flutter instability in the forced response of the lattice and its corresponding effective medium.

\section{A premise on lack of symmetry and the energy absorbed or released in an infinitesimal strain cycle}
\label{sec:energy_loop}
The crucial feature of the elastic materials explored in this article is that the lack of the major symmetry in the constitutive fourth-order tensor leads a material to absorb or release energy in a closed infinitesimal cycle of strain, when this cycle encloses a non-vanishing area.
Usually this feature is believed to be obvious, so that proofs of it are never reported, even in elementary or specialized textbooks.
Therefore, the precise link between the violation of conservation of strain energy in a material and the non-symmetry of its acoustic tensor is detailed below.

The effective material that will be shown to correspond to the incremental response of a lattice prestressed with follower forces is characterized by an incremental response, where, adopting an updated Lagrangian description, the increment of the first Piola-Kirchhoff stress, $\dot{\bS}$, is linearly related to the gradient of incremental displacement $\bL$ through the fourth-order elastic tensor $\fH$ as
\begin{equation}
    \label{eq:constitutive_law}
    \dot{\bS} = \fH[\bL] \,.
\end{equation}
Being the increments in the first Piola-Kirchhoff stress and in the displacement gradient both unsymmetric, the fourth-order tensor $\fH$ lacks the left and right minor symmetries, respectively, $\fH_{ijhk} \neq \fH_{jihk}$ and $\fH_{ijhk} \neq \fH_{ijkh}$.
When an incremental strain potential exists, the tensor $\fH$ possesses the major symmetry, $\fH_{ijhk}=\fH_{hkij}$.
In the case the response is not hyper-elastic, tensor $\fH$ lacks the major symmetry, $\fH_{ijhk} \neq \fH_{hkij}$, which is the case addressed in the present article.

Body-wave propagation in a material incrementally defined through Eq.~\eqref{eq:constitutive_law} is governed by the acoustic tensor $\bA$, function of the unit vector $\bn$, which determines the propagation direction, and defined for every vector $\bg$ as
\begin{equation}
    \label{eq:acoustic_tensor}
    \bA(\bn) \bg = \fH[\bg \otimes \bn]\bn \,.
\end{equation}

Note that lack of major symmetry of $\fH$ is equivalent to lack of symmetry of $\bA$, so that when a strain potential does not exist, $\bA$ is unsymmetric.

The response of a material is stable when all eigenvalues of the acoustic tensor are real and positive.
Loss of ellipticity is defined as the vanishing of one of the  eigenvalues and corresponds to the possibility of localization of deformation and shear banding.
When two eigenvalues of the acoustic tensor are complex conjugate either flutter instability (the real part of the eigenvalues is different from zero) or divergence instability (the eigenvalues are a pure imaginary conjugate pair) occurs.

In order to address the consequences of the lack of the major symmetry of the fourth-order tensor $\fH$, it is instrumental to define the incremental stress $\dot{\bS}$ generated by an incremental displacement gradient assumed in the form $\bL = \bg\otimes\bn$, which is
\begin{equation*}
    \dot{\bS} = \fH[\bg \otimes \bn] \,.
\end{equation*}
The infinitesimal \textit{second-order work} done by the incremental stress in an infinitesimal increment of strain in which $\bn$ remains constant and $\bg$ is incremented by $d\bg$ can be written as
\begin{equation*}
    d\mW = \dot{\bS} \scalp d\bL = \fH[\bg \otimes \bn] \scalp d\bg \otimes \bn = d\bg \scalp \bA(\bn) \bg \,,
\end{equation*}
where $\bA(\bn)$ is the acoustic tensor, Eq.~\eqref{eq:acoustic_tensor}.
\begin{figure}[htb]
    \centering
    \includegraphics[width=0.33\linewidth]{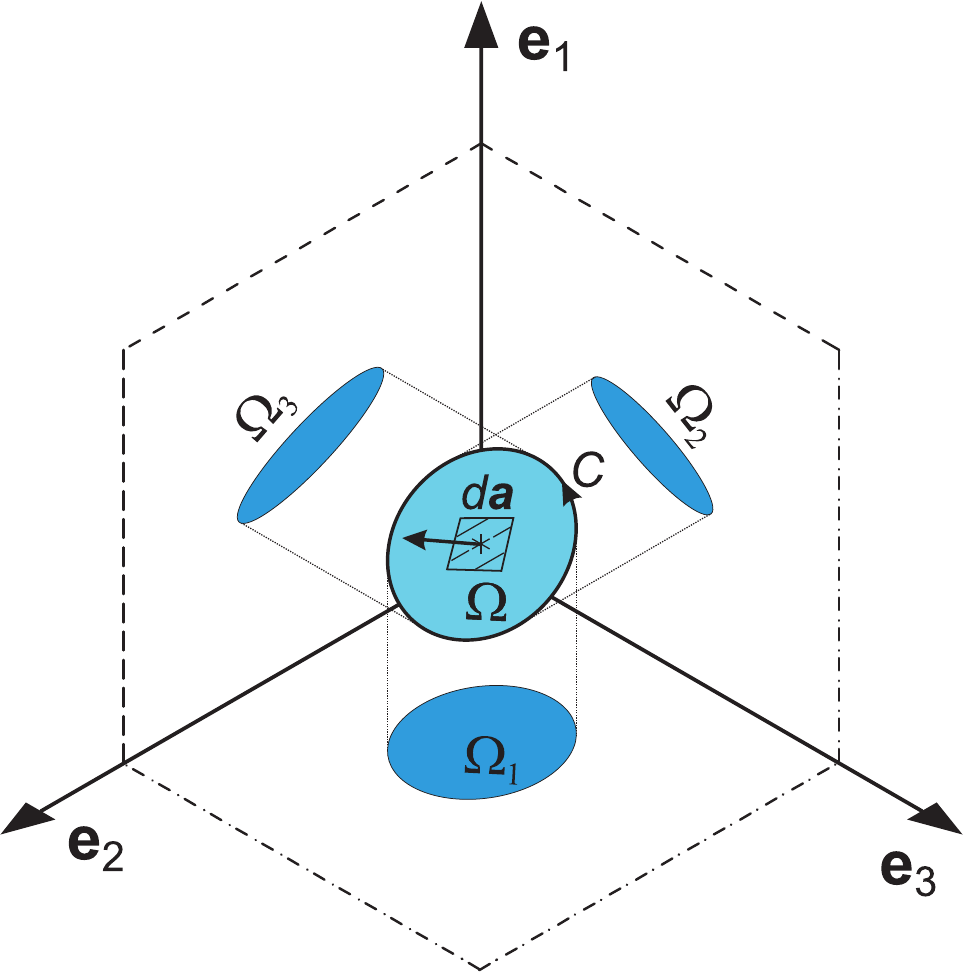}
    \caption{
        Representative sketch of an incremental strain loop in the $\bg$-space with enclosed area $\Omega$ and `signed' projections $\Omega_1$, $\Omega_2$, $\Omega_3$ involved in the energy absorbed by a non-hyper-elastic material in the closed cycle $C$, Eq.~\eqref{eq:work_closed_cycle_result}.
    }
    \label{fig:loop_projections}
\end{figure}
Thus, the work done in a \textit{closed} infinitesimal loop $C$ in the three-dimensional space characterized by the orthonormal basis $\{\be_1, \be_2, \be_3\}$, where the components of $\bg$ are $g_1$, $g_2$, and $g_3$ (Fig.~\ref{fig:loop_projections}), can be written as
\begin{equation}
    \label{eq:work_closed_cycle}
    \mW(C) = \oint_C \left( \bA\bg \right) \scalp d\bg \,,
\end{equation}
where the dependence of $\bA$ on $\bn$ is omitted.
An application of the Stokes theorem to Eq.~\eqref{eq:work_closed_cycle} provides
\begin{equation}
    \label{eq:stokes}
    \oint_C \left( \bA\bg \right)\scalp d\bg = \int_\Omega \curl\left(\bA\bg\right) \scalp d\ba \,,
\end{equation}
where the $\curl$ is performed with respect to $\bg$ and $\Omega$ is an open surface bounded by $C$ and having $d\ba$ as area element, oriented according to the right-hand rule.

The $\curl$ on the right-hand side of Eq.~\eqref{eq:stokes} is constant with respect to $\bg$ and assumes the form
\begin{equation}
    \label{eq:curl_acoustic_tensor}
    \curl( \bA \bg ) = 2A_{[32]} \be_1+ 2A_{[13]} \be_2 + 2A_{[21]} \be_3 \,,
\end{equation}
where $A_{[ij]} = (A_{ij}-A_{ji})/2$ is the skew-symmetric part of $\bA$.
Therefore, \textit{only the skew-symmetric part of $\bA$ plays a role in determining the work done along a closed contour $C$} in the strain space, so that a chain substitution of Eqs.~\eqref{eq:curl_acoustic_tensor},~\eqref{eq:stokes} into Eq.~\eqref{eq:work_closed_cycle} yields
\begin{equation}
    \label{eq:work_closed_cycle_result}
    \mW(C) = 2A_{[32]} \underbrace{\int_\Omega \be_1\scalp d\ba}_{\Omega_1} + 2A_{[13]} \underbrace{\int_\Omega \be_2\scalp d\ba}_{\Omega_2} + 2A_{[21]} \underbrace{\int_\Omega \be_3\scalp d\ba}_{\Omega_3} \,,
\end{equation}
where $\Omega_1$, $\Omega_2$, $\Omega_3$ can be interpreted as the `signed' areas of the projections of the surface $\Omega$ onto the coordinate planes (of unit normal $\be_i$, see Fig.~\ref{fig:loop_projections}).

The significance of Eq.~\eqref{eq:work_closed_cycle_result} lies in the fact that it provides a direct way for the determination of the loop that maximizes the absorbed/released energy.
Specifically, a loop whose area elements are aligned parallel to and in the same direction of (in the opposite direction of) the $\curl$ vector~\eqref{eq:curl_acoustic_tensor} will maximizes the scalar product~\eqref{eq:work_closed_cycle_result} and thus the energy absorbed (released).
In other words, the energy~\eqref{eq:work_closed_cycle_result} is maximized/minimized by loops contained in the plane \textit{orthogonal} to the $\curl$ vector~\eqref{eq:curl_acoustic_tensor}.

For the two-dimensional setting adopted throughout the rest of the article, Eq.~\eqref{eq:work_closed_cycle_result} is reduced to
\begin{equation}
    \label{eq:work_closed_cycle_result_2d}
    \mW(C) = \left( A_{21} - A_{12} \right) \Omega_3 \,,
\end{equation}
where $\Omega_3$ is the `signed' area of the two-dimensional loop $C$, positive (negative) for counterclockwise (clockwise) loops.
Eq.~\eqref{eq:work_closed_cycle_result_2d} proves that the material absorbs (releases) a net quantity of energy along \textit{every} counterclockwise cycle in the $\bg$-space for which the skew-symmetric term $A_{21} - A_{12}$ is positive (negative).

The actual realization of an effective material possessing an unsymmetric acoustic tensor is the central subject of the following sections.

\section{Incremental dynamics of lattices preloaded by follower forces}
\label{sec:problem_setting}
%
%
A way is presented to introduce non-selfadjointness in the mechanics of periodic lattices, via non-conservative loadings.
In particular, it is shown how \textit{a suitable periodic configuration of follower loads applied to a lattice of elastic rods (deforming in a plane under bending, axial and shear forces) may induce a prestress state leading to a non-Hermitian incremental response of the system}.

Note that the follower loads are transmitted by the `environment' on the structure and are assumed to be non-conservative of the tangential type, as in the well-known case of the Ziegler double pendulum.
Their practical realization cannot be achieved using tools based on a potential energy (such as for instance springs), but necessitate special devices such as those developed by Bigoni and co-workers~\cite{bigoni_2011,bigoni_2018a,cazzolli_2020}.

\begin{figure}[htb]
    \centering
    \includegraphics[width=\linewidth]{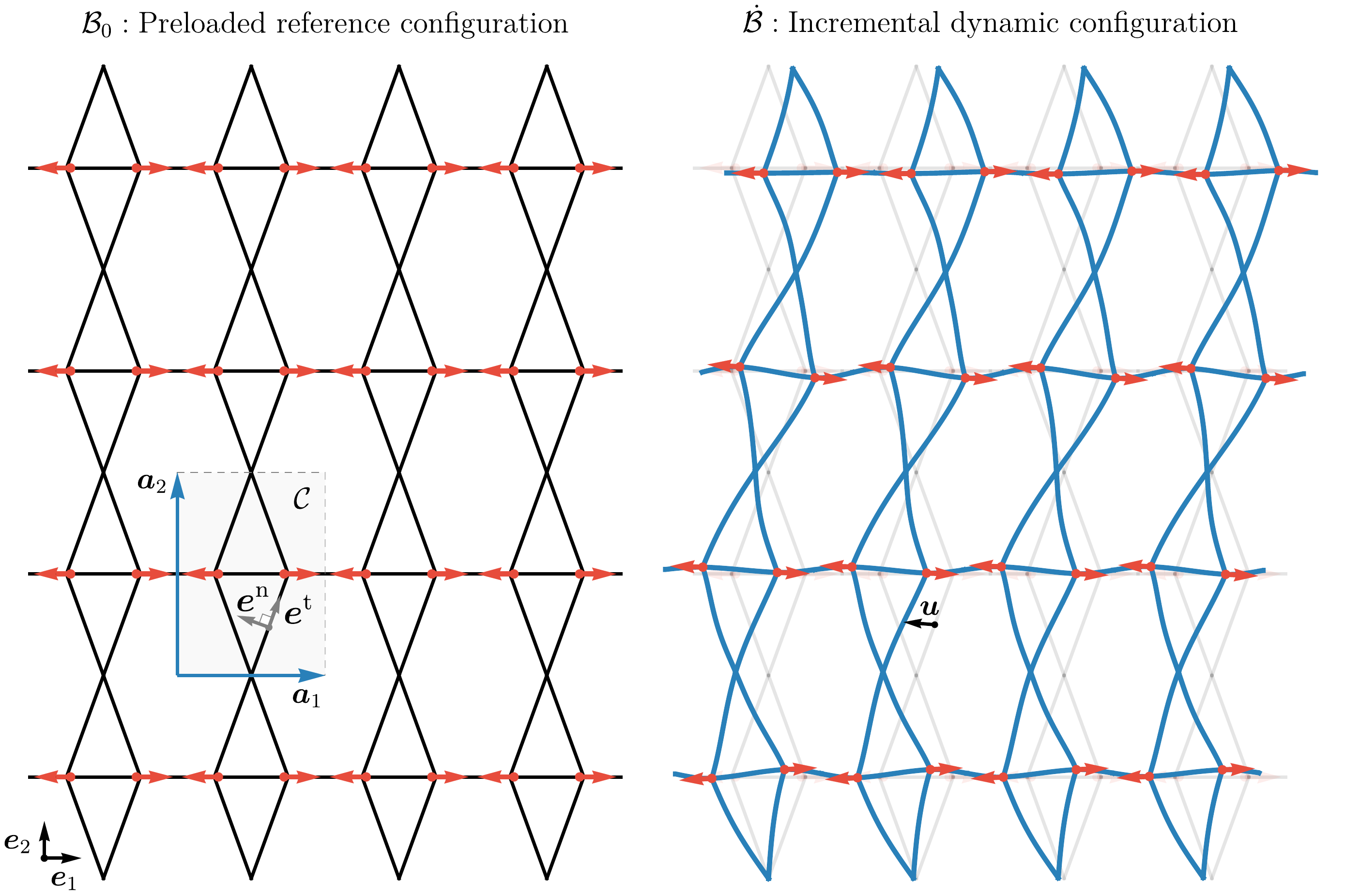}
    \caption{
        Two-dimensional lattice structure periodically preloaded by follower forces.
        Periodicity of the system is defined by the unit cell $\mC$ and the direct basis $\{\ba_1,\ba_2\}$.
        The preloaded configuration $\mB_0$ (shown on the left) is assumed as reference for the incremental displacement field sketched on the right.
        Homogenization shows that the effective incremental behavior of the lattice is described by an elastic material which does \textit{not} follow from a scalar strain potential.
        See also Fig.~\ref{fig:problem_setting_detailed} for a detailed view of the unit cell.
    }
    \label{fig:problem_setting}
\end{figure}

\begin{figure}[htbp]
    \centering
    \includegraphics[width=0.95\linewidth]{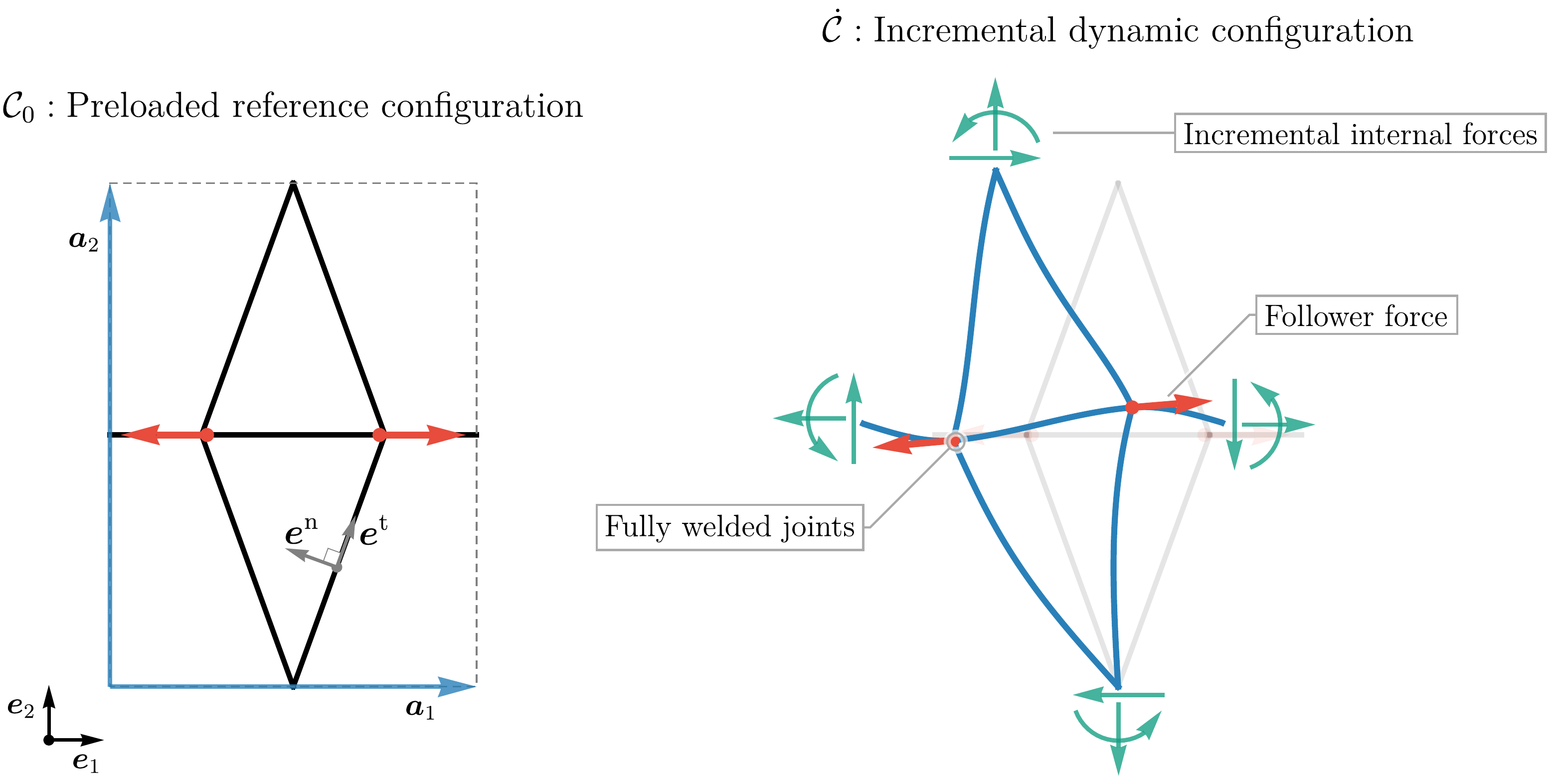}
    \caption{
        Detailed view of the unit cell $\mC$ of the lattice structure illustrated in Fig.~\ref{fig:problem_setting}.
        All joints connecting the rods are assumed to be fully rigid, so that axial and shear forces and bending moments are incrementally generated.
        The follower forces, sketched in red, are non-conservative and remain tangential to the deformed (horizontal) rod.
    }
    \label{fig:problem_setting_detailed}
\end{figure}

An infinite grid (with nodes preserving continuity of displacements and rotations, so that incremental bending moment is transmitted in addition to shear and normal forces) of elastic rods is considered, obtained through the tessellation of a single unit cell $\mC$ along two linearly independent directions, defined by a pair of basis vectors $\{\ba_1,\ba_2\}$, Figs.~\ref{fig:problem_setting} and~\ref{fig:problem_setting_detailed}.
This configuration is assumed to satisfy equilibrium with the externally applied forces and subject to a periodic state of axial stress inside the rods, the \textit{prestress}.
The incremental displacement along the rods is defined by the vector field
\begin{equation}
    \label{eq:displacement_field}
    \bu_k(s_k,t) = u_k(s_k,t)\,\be^{\text{t}}_k + v_k(s_k,t)\,\be^{\text{n}}_k \,, \quad \forall k\in\{1,\dots,N_{\text{b}}\} \,,
\end{equation}
where $\be^{\text{t}}_k$ and $\be^{\text{n}}_k$ are unit vectors aligned parallel and orthogonal to the rods (Figs.~\ref{fig:problem_setting} and~\ref{fig:problem_setting_detailed}), so that $u_k$ and $v_k$ denote the axial and transverse displacements, functions of the local coordinate $s_k$ and of time, while $N_{\text{b}}$ is the number of rods in a unit cell.
As the rods are assumed unshearable, the incremental rotation of the cross section is defined as $\theta_k(s_k,t)=v_k'(s_k,t)$.

The linearized equations of motion governing the incremental displacement fields, to be superimposed on a given axially-preloaded configuration of a rod (the index $k$ is momentarily omitted), are
\begin{subequations}
    \label{eq:equations_motion_beam}
    \begin{gather}
        \label{eq:equations_motion_beam_u}
        \gamma\,\ddot{u}(s,t) - A\,u''(s,t) = 0 \,, \\
        \label{eq:equations_motion_beam_v}
        \gamma\,\ddot{v}(s,t) + B\,v''''(s,t) - P\,v''(s,t) = 0 \,,
    \end{gather}
\end{subequations}
where $A$ and $B$ are the axial and bending stiffness, respectively, the intertia is defined by the linear mass density $\gamma$, and $P$ is the axial preload, generated by external loads of the tangentially follower type (the inclusion of additional dead loadings is trivial and will not be considered for simplicity).

For \textit{time-harmonic} motion, $\bu(s,t)=\bu(s)\,e^{-i\,\omega t}$, the solution to Eqs.~\eqref{eq:equations_motion_beam} can be cast in the following form
\begin{equation}
    \label{eq:shape_functions}
    \bu(s) = \bN(s; \omega^2, P)\,\bq \,,
\end{equation}
where vector $\bq$ collects the generalized \textit{incremental displacements} (including rotations) at the ends of the rod and $\bN$ is a matrix of shape functions defining the exact solution.
The \textit{incremental forces} transmitted at the ends of the rod can be evaluated via Eq.~\eqref{eq:shape_functions} and the second-order kinetic and elastic potential energies for the $k$-th rod, denoted here as $\mT_k$ and $\mV_k$, can be represented as~\cite{bordiga_2021}
\begin{equation}
    \label{eq:kinetic_potential_energy_beam}
    \mT_k(\dot{\bq}_k) = \frac{1}{2}\,\dot{\bq}_k \scalp \bM_k(\omega^2,P_k)\,\dot{\bq}_k \,, \quad
    \mV_k(\bq_k) = \frac{1}{2}\,\bq_k \scalp \bK_k(\omega^2,P_k)\,\bq_k \,, \quad \forall k\in\{1,\dots,N_{\text{b}}\} \,,
\end{equation}
where $\bM_k$ and $\bK_k$ are the \textit{symmetric} mass and stiffness matrices of the $k$-th rod.

The second-order functionals~\eqref{eq:kinetic_potential_energy_beam} can conveniently be employed to build the Lagrangian of the unit cell, $\mL(\bq,\dot{\bq}) = \mT(\dot{\bq}) - \mV(\bq)$, by simply summing the energy contribution of each rod,
\begin{equation}
    \label{eq:kinetic_potential_energy_cell}
    \mT(\bq,\dot{\bq}) = \sum_{k=1}^{N_{\text{b}}} \mT_k(\dot{\bq}_k) \,, \quad \mV(\bq) = \sum_{k=1}^{N_{\text{b}}} \mV_k(\bq_k) \,,
\end{equation}
where vector $\bq$ collects all generalized \textit{incremental} displacements of the unit cell.
Therefore, the incremental dynamics of the unit cell is governed by the variational principle
\begin{equation}
    \label{eq:Hamilton_principle_cell}
    -\delta \int_{t_0}^{t_1} \mL(\bq,\dot{\bq})\,\text{d}t = \int_{t_0}^{t_1} \bef\scalp\delta\bq\,\text{d}t \,, \quad \forall \delta\bq\,\lvert\,\delta\bq(t_0)=\delta\bq(t_1)=0\,,
\end{equation}
where $\bef\scalp\delta\bq$ is the virtual work done by the generalized \textit{incremental} forces acting on the nodes of the unit cell.
The use of Eq.~\eqref{eq:kinetic_potential_energy_cell} in Eq.~\eqref{eq:Hamilton_principle_cell} yields the equations of motion in the Lagrangian form
\begin{equation}
    \label{eq:equations_of_motion_cell_Lagrangian}
    \frac{\text{d}}{\text{d}t}\deriv{\mL(\bq,\dot{\bq})}{\dot{\bq}} - \deriv{\mL(\bq,\dot{\bq})}{\bq} = \bef \,,
\end{equation}
which more explicitly becomes
\begin{equation}
    \label{eq:equations_of_motion_cell}
    \bM(\omega^2,\bP)\,\ddot{\bq} + \bK_{\text{H}}(\omega^2,\bP)\,\bq = \bef \,,
\end{equation}
where $\bP$ is a $N_{\text{b}}$-dimensional vector collecting the values of preload of the rods, while $\bM$ and $\bK_{\text{H}}$ are the \textit{symmetric} mass and stiffness matrix of the unit cell.

It is important to remark that the derivation of Eqs.~\eqref{eq:Hamilton_principle_cell}--\eqref{eq:equations_of_motion_cell} does not involve any restrictive assumption on the nature of the incremental forces $\bef$.
If these forces are derived from a potential, a complete Hermitian formulation follows and homogenization leads to a hyper-elastic effective material.
Therefore, the way to hypo-elasticity is to assume forces not descending from a potential and, in particular, \textit{follower} forces will be assumed, remaining constant in modulus and tangent to the structural element at the point where these are applied.

\subsection{Incremental virtual work contribution of follower forces}
\label{sec:follower_contribution}
%
%
The incremental force vector $\bef$ on the right-hand side of Eqs.~\eqref{eq:Hamilton_principle_cell}--\eqref{eq:equations_of_motion_cell} is conveniently split as $\bef=\hat{\bef}+\bef^{\text{F}}$, with $\bef^{\text{F}}$ denoting the incremental contribution of the follower loads (occurring as a consequence of an incremental deformation), assumed to be applied only to the \textit{inner} nodes of the unit cell, and $\hat{\bef}$ being the forces internal to the lattice that are transmitted by the neighboring cells.
Thus $\bef^{\text{F}}$ has non-zero entries corresponding to inner nodes only and $\hat{\bef}$ has non-zero entries corresponding to boundary nodes only.
The incremental force vector $\bef^{\text{F}}$ due to follower loads is derived as follows.
\begin{figure}[htb]
    \centering
    \includegraphics[width=0.8\linewidth]{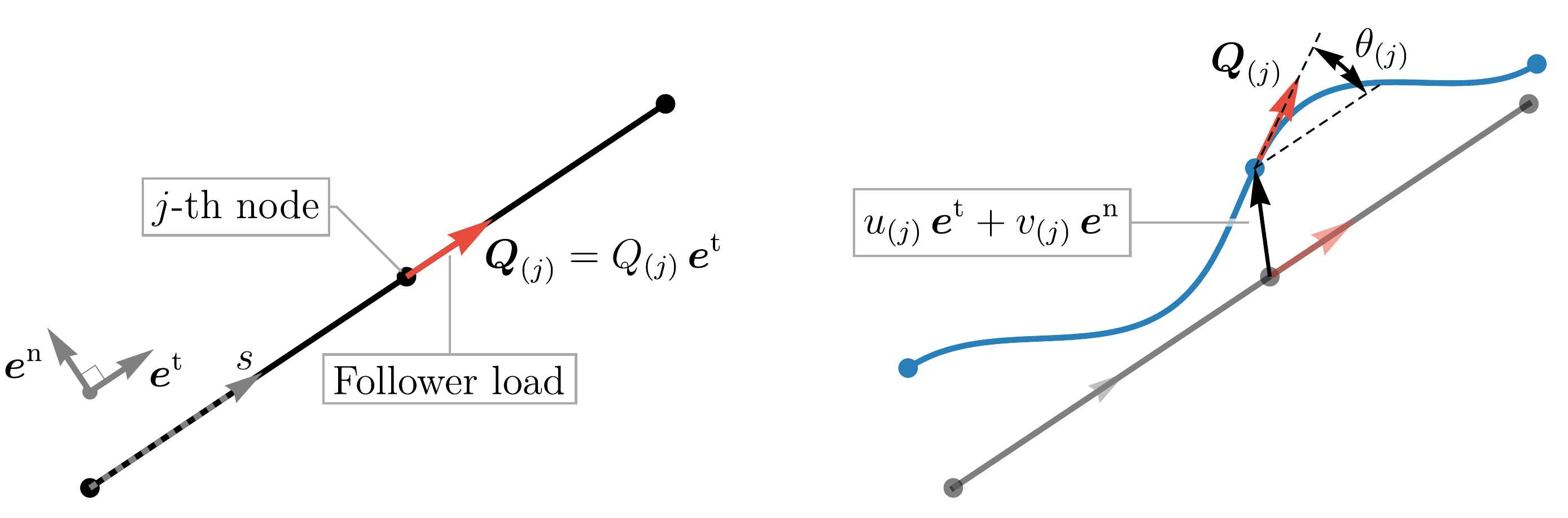}
    \caption{
        Two rods of the lattice loaded by a follower force of constant magnitude and remaining always tangential to the rods at the location where it is applied.
        The reference configuration is shown on the left, while an incrementally deformed configuration is sketched on the right.
        Note that the application of the concentrated follower force generates an axial preload inside the rods that is discontinuous at the junction node $j$ but constant on each rod.
    }
    \label{fig:follower_force_rods}
\end{figure}

With reference to Fig.~\ref{fig:follower_force_rods}, a pair of rods of the lattice is considered loaded by a tangential follower force $\bQ_{(j)}$, applied to the junction node $j$.
For convenience, the generalized incremental displacement vector of the $j$-th node is denoted as $\bq_{(j)}=u_{(j)}\,\be^{\text{t}}+v_{(j)}\,\be^{\text{n}}+\theta_{(j)}\,\be^{\text{3}}$, with $\be^{\text{3}}$ being the out-of-plane unit vector $\be^{\text{3}}=\be^{\text{t}}\times\be^{\text{n}}$.
Correspondingly, the generalized incremental force vector acting at the $j$-th node assumes the form $\bef_{(j)} = H_{(j)}\,\be^{\text{t}}+V_{(j)}\,\be^{\text{n}}+M_{(j)}\,\be^{\text{3}}$, where $H_{(j)}$ is the axial load, $V_{(j)}$ the transverse load and $M_{(j)}$ the out-of-plane bending moment.
With this notation, the linearization of the virtual work associated to the follower force, performed with respect to the preloaded equilibrium configuration, reads as
\begin{equation}
    \label{eq:linearized_follower_virtual_work}
    Q_{(j)} \left(\cos\theta_{(j)}\,\be^{\text{t}} + \sin\theta_{(j)}\,\be^{\text{n}}\right) \scalp \delta\bq_{(j)} \sim
    Q_{(j)}\,\left(\be^{\text{t}} + \theta_{(j)}\,\be^{\text{n}}\right) \scalp \delta\bq_{(j)} \,,
\end{equation}
where $Q_{(j)}=\norm{\bQ_{(j)}}$ and $\theta_{(j)}$ is the incremental rotation of the node.
As the linearization is made about an equilibrium configuration, the zeroth-order term $Q_{(j)}\,\be^{\text{t}} \scalp \delta\bq_{(j)}$ is assumed to be balanced by the internal axial loads $\bP$, so that only the difference $Q_{(j)}\,\left(\be^{\text{t}} + \theta_{(j)}\,\be^{\text{n}}\right) \scalp \delta\bq_{(j)} - Q_{(j)}\,\be^{\text{t}} \scalp \delta\bq_{(j)}$ contributes to the incremental problem, Eq.~\eqref{eq:equations_of_motion_cell}.
Hence, the follower load performs the incremental virtual work
\begin{equation}
    \label{eq:incremental_work_follower}
    \bef^{\text{F}}_{(j)} \scalp \delta\bq_{(j)} = Q_{(j)}\,\theta_{(j)}\,\be^{\text{n}} \scalp \delta\bq_{(j)} \,.
\end{equation}
Eq.~\eqref{eq:incremental_work_follower} shows that the \textit{increment} of the follower force is always \textit{orthogonal} to the follower force itself, so that the incremental force $\bef^{\text{F}}_{(j)}$ can be expressed by means of the vectorial relation $\be^{\text{n}}=\be^3\times\be^{\text{t}}$ as
\begin{equation}
    \label{eq:increment_follower_force}
    \bef^{\text{F}}_{(j)} = Q_{(j)}\,\theta_{(j)}\,\be^{\text{n}} = \left(\theta_{(j)}\,\be^3\right) \times \bQ_{(j)} \,.
\end{equation}
Furthermore, Eq.~\eqref{eq:incremental_work_follower} can be rewritten more explicitly as
\begin{equation}
    \label{eq:incremental_work_follower_current}
    \bef^{\text{F}}_{(j)} \scalp \delta\bq_{(j)} = Q_{(j)}\, \theta_{(j)} \, \delta v_{(j)} \,,
\end{equation}
an expression that highlights the \textit{unsymmetry of the follower contribution}, as it is clear that no scalar function $\phi$ exists such that $\delta\phi=Q_{(j)}\, \theta_{(j)} \, \delta v_{(j)}$.

It may be worth noting that, due to the external concentrated (follower) loading, the state of preload in the lattice is necessarily piecewise constant.
Consequently, the energies, Eq.~\eqref{eq:kinetic_potential_energy_beam}, have to be evaluated for each rod, considering the local value of preload.
Moreover, the displacement vector of the unit cell $\bq$ necessarily contains the displacements at all the loading points.

The incremental contribution from the follower force, Eq.~\eqref{eq:incremental_work_follower}, to the right-hand side of the variational principle, Eq.~\eqref{eq:Hamilton_principle_cell}, can be computed by summing the contributions from all nodes where follower loads are applied.
Hence, the total incremental virtual work associated with the follower loads can be expressed using Eq.~\eqref{eq:incremental_work_follower} as
\begin{equation}
    \label{eq:total_incremental_work_follower}
    \bef^{\text{F}} \scalp \delta\bq = \sum_{j=1}^N \bef^{\text{F}}_{(j)} \scalp \delta\bq_{(j)} = \sum_{j=1}^N \left(\theta_{(j)}\,\be^3\right) \times \bQ_{(j)} \scalp \delta\bq_{(j)} \,,
\end{equation}
where $N$ is the number of nodes in the unit cell.

Eq.~\eqref{eq:total_incremental_work_follower} uniquely defines the incremental force vector $\bef^{\text{F}}$ as a \textit{linear} function of the nodes' incremental rotation $\theta_{(j)}$.
As a consequence, a unique matrix $\bK_{\text{F}}$, which will be called \textit{follower load matrix}, can be defined
\begin{equation}
    \label{eq:follower_matrix}
    \bef^{\text{F}}(\bq) = -\bK_{\text{F}}(\bP)\,\bq \,,
\end{equation}
where it is clear that $\bK_{\text{F}}\neq\trans{\bK_{\text{F}}}$ due to the fact that the virtual work contribution of each follower load is unsymmetric with respect to the exchange between rotation and transverse displacement, as shown by Eq.~\eqref{eq:incremental_work_follower_current}.
Note that the state of preload $\bP$ in the reference configuration is assumed to be in equilibrium with the externally applied follower forces $\bQ_{(j)}$, and therefore $\bK_{\text{F}}$ depends linearly on the preload state $\bP$.

Through the use of Eq.~\eqref{eq:follower_matrix}, the equations of motion of the unit cell~\eqref{eq:equations_of_motion_cell} can be cast in the following form
\begin{equation}
    \label{eq:equations_of_motion_cell_follower}
    \bM(\omega^2,\bP)\,\ddot{\bq} + \left( \bK_{\text{H}}(\omega^2,\bP) + \bK_{\text{F}}(\bP) \right)\bq = \hat{\bef} \,,
\end{equation}
an equation clearly showing that the presence of follower loads provides the additional \textit{unsymmetric} stiffness term $\bK_{\text{F}}$ and also affects the matrices $\bM$ and $\bK_{\text{H}}$ through the preload $\bP$.
It should be noted that if $\bK_{\text{F}}$ is set to zero with $\bP\neq\bzero$, Eq.~\eqref{eq:equations_of_motion_cell_follower} reduces to the governing equation holding for the same structure subject to the same loads, but now assumed dead.

\subsection{Wave propagation in non-Hermitian lattices}
\label{sec:wave_propagation}
%
%
The problem of incremental wave propagation can now be formulated for the infinite lattice structure preloaded by the periodic distribution of follower forces.

By enforcing Floquet-Bloch boundary conditions in the usual linear form $\bq=\bZ(\bk)\,\tilde{\bq}$, with $\bk\in\Reals^3$ being the wave vector and $\tilde{\bq}$ the reduced vector of degrees of freedom and recalling that $\conjtrans{\bZ(\bk)}\hat{\bef}=\bzero$ (where the superscript $\conjtrans{}$ denotes Hermitian conjugation)~\cite{phani_2006,bordiga_2021}, Eq.~\eqref{eq:equations_of_motion_cell_follower} becomes
\begin{equation}
    \label{eq:eigenvalue_problem}
    \conjtrans{\bZ(\bk)} \left( -\omega^2 \bM(\omega^2,\bP) + \bK(\omega^2,\bP) \right) \bZ(\bk)\,\tilde{\bq} = \bzero \,,
\end{equation}
where $\bK(\omega^2,\bP)=\bK_{\text{H}}(\omega^2,\bP)+\bK_{\text{F}}(\bP)$ is the total \textit{unsymmetric} stiffness matrix.
The system~\eqref{eq:eigenvalue_problem} can concisely be rewritten as
\begin{equation}
    \label{eq:eigenvalue_problem_compact}
    \bA(\omega^2,\bk,\bP)\,\tilde{\bq} = \bzero \,,
\end{equation}
upon introduction of the matrix
\begin{equation}
    \label{eq:system_matrix}
    \bA(\omega^2,\bk,\bP) = \conjtrans{\bZ(\bk)} \left( -\omega^2 \bM(\omega^2,\bP) + \bK(\omega^2,\bP) \right) \bZ(\bk) \,.
\end{equation}

Eq.~\eqref{eq:eigenvalue_problem_compact} governs the incremental wave propagation for the infinite lattice and has the structure of an eigenvalue problem for the eigenfrequencies $\omega$ and eigenvectors $\tilde{\bq}$.
It is highlighted that the \textit{non-selfadjointness is related to the prestress-dependence generated through follower forces}, corresponding to a non-vanishing and unsymmetric stiffness contribution $\bK_{\text{F}}$, so that
\begin{equation*}
    \conjtrans{\bA(\omega^2,\bk,\bP)} = \conjtrans{\bZ(\bk)} \left( -\conj{\omega}^2 \bM(\conj{\omega}^2,\bP) + \bK(\conj{\omega}^2,\bP) + \trans{\bK_{\text{F}}(\bP)} \right) \bZ(\bk) \neq \bA(\conj{\omega}^2,\bk,\bP) \,.
\end{equation*}
It is therefore clear that when the follower contribution vanishes, either because only dead loads are prescribed or because the lattice is not preloaded, Eq.~\eqref{eq:eigenvalue_problem} becomes a standard eigenvalue problem for wave propagation in a \textit{conservative} structure with \textit{real} eigenvalues, $\omega^2\in\Reals$.

\textit{The presence of the follower loads alters drastically the stability properties of the lattice material}.
In fact, while for dead loads ($\bK_{\text{F}}=\bzero$, $\bP\neq\bzero$) the governing matrix $\bA(\omega^2)$ is Hermitian\footnote{
A matrix--valued function $\bA(\lambda)$ is said to be Hermitian if $\bA(\lambda)^{\text{H}} = \bA(\conj{\lambda})$ for all $\lambda \in \Complexes$.
} and the stability is defined in the usual way as
\begin{equation}
    \label{eq:stability_no_follower}
    \conjtrans{\bA(\omega^2)}=\bA(\conj{\omega}^2) \, \Longrightarrow \, \omega^2\in\Reals \,
    \begin{cases}
        \omega^2>0    & \text{stability}              \\
        \omega^2\leq0 & \text{divergence instability} \\
    \end{cases}
\end{equation}
when \textit{follower loads are present ($\bK_{\textup{F}}\neq\bzero$, $\bP\neq\bzero$), the governing matrix $\bA(\omega^2)$ is non-Hermitian and the stability scenario completely changes}
\begin{equation}
    \label{eq:stability_follower}
    \conjtrans{\bA(\omega^2)}\neq\bA(\conj{\omega}^2) \, \Longrightarrow \, \omega^2\in\Complexes \,
    \begin{cases}
        \Ip(\omega^2)=0     & \text{stability}\, (\omega^2>0)       \\
                            & \text{or divergence}\,(\omega^2\leq0) \\
        \Ip(\omega^2)\neq 0 & \text{flutter instability}            \\
    \end{cases}
\end{equation}
and \textit{flutter instability becomes possible}.

\section{Non-hyper-elastic effective continuum via non-Hermitian lattice wave asymptotics}
\label{sec:homogenization}
%
%
A low-frequency and long-wavelength (LF-LW) asymptotic analysis of the lattice waves is developed to derive the \textit{macroscopic} response of the lattice structure from the dynamics governed by Eq.~\eqref{eq:eigenvalue_problem_compact}.
This technique, never so far applied to non-Hermitian systems, shows that the unsymmetry of the problem~\eqref{eq:eigenvalue_problem} leads to a fundamentally different asymptotics compared to the classical homogenization of Hermitian problems~\cite{bordiga_2021}, so that non-trivial solvability conditions come into play.

The asymptotic approach provides a positive answer to the fundamental question of whether \textit{the non-selfadjointness of the wave propagation problem leads to a non-selfadjoint macroscopic response}.
Hence, \textit{the effective continuum is necessarily non-hyper-elastic}, for which a strain-energy function does not exists, even though the macroscopic incremental response remains both linear and elastic.

\subsection{Lattice wave asymptotics}
\label{sec:lattice_wave_asymptotics}
%
%
The LF-LW asymptotics of lattice waves is developed by assuming the origin of the $\{\omega,\bk\}$-space as center for a series expansion of the matrix $\bA(\omega(\bk)^2,\bk)$ and its right eigenvector $\tilde{\bq}(\bk)$.
Accordingly, by setting $\bk=\epsilon\,\bh$ with $\epsilon$ being a small dimensionless parameter and $\bh$ a fixed vector in the $\bk$-space, the eigenvalue $\omega(\epsilon\bh)$ and the eigenvector $\tilde{\bq}(\epsilon\bh)$ are sought in a series form
\begin{subequations}
    \label{eq:omega_waveform_expansion}
    \begin{align}
        \label{eq:omega_expansion}
        \omega(\epsilon\bh)      & \sim \omega_{\bh}^{(1)}\,\epsilon + \omega_{\bh}^{(2)}\,\epsilon^2 + \dots \,,                                     \\
        \label{eq:waveform_expansion}
        \tilde{\bq}(\epsilon\bh) & \sim \tilde{\bq}_{\bh}^{(0)} + \tilde{\bq}^{(1)}_{\bh}\,\epsilon + \tilde{\bq}^{(2)}_{\bh}\,\epsilon^2 + \dots \,,
    \end{align}
\end{subequations}
as solutions of the corresponding expansion of the system~\eqref{eq:eigenvalue_problem_compact}, which becomes
\begin{equation}
    \label{eq:system_expansion}
    \left(\bA^{(0)} + \bA^{(1)}_{\bh}\,\epsilon + \bA^{(2)}_{\bh}\,\epsilon^2 + \dots \right)\left( \tilde{\bq}_{\bh}^{(0)} + \tilde{\bq}^{(1)}_{\bh}\,\epsilon + \tilde{\bq}^{(2)}_{\bh}\,\epsilon^2 + \dots \right) = \bzero \,,
\end{equation}
where the subscript $_{\bh}$ is meant to remind the dependence on the wave vector along which the expansion is performed.
As Eq.~\eqref{eq:system_expansion} has to be satisfied at each order, a hierarchy of linear systems is obtained, which, up to second-order, reads
\begin{equation}
    \label{eq:system_sequence}
    \begin{aligned}
         & \mO(\epsilon^0): \quad \bA^{(0)}\,\tilde{\bq}^{(0)}_{\bh} = \bzero \,,                                                                                       \\
         & \mO(\epsilon^1): \quad \bA^{(0)}\,\tilde{\bq}^{(1)}_{\bh} + \bA^{(1)}_{\bh}\,\tilde{\bq}^{(0)}_{\bh} = \bzero \,,                                            \\
         & \mO(\epsilon^2): \quad \bA^{(0)}\,\tilde{\bq}^{(2)}_{\bh} + \bA^{(1)}_{\bh}\,\tilde{\bq}^{(1)}_{\bh} + \bA^{(2)}_{\bh}\,\tilde{\bq}^{(0)}_{\bh} = \bzero \,.
    \end{aligned}
\end{equation}

The matrix $\bA^{(0)}$ is real and \textit{singular} with a non-trivial nullspace spanned by rigid-body translations, as so-called \textit{floppy modes}~\cite{mao_2018,zhang_2018,ma_2018} are excluded from the present analysis~\cite{bordiga_2021}.
Consequently, the existence of solutions for Eqs.~\eqref{eq:system_sequence} has to be assessed through the Fredholm alternative theorem.
The solvability condition for the $j$-th equation of the system~\eqref{eq:system_sequence} can thus be written as follows
\begin{equation}
    \label{eq:solvability_condition}
    \by \scalp \sum_{p=1}^{j} \bA^{(p)}_{\bh}\,\bq^{(j-p)}_{\bh} = 0 \,, \quad \forall \by\in\ker\trans{{\bA^{(0)}}} \,, \quad \forall j>0 \,.
\end{equation}
where the space spanned by the vectors $\by$ is the so-called left-nullspace of $\bA^{(0)}$, in other words, the space of left eigenvectors $\by$ satisfying
$\trans{{\bA^{(0)}}}\by=\bzero$.
Note that, the Rouché--Capelli theorem implies that the nullspace and the left-nullspace of the square matrix $\bA^{(0)}$ must have the same dimension.

The crucial point to be highlighted here is that for the non-Hermitian problem~\eqref{eq:eigenvalue_problem_compact}, the matrix $\bA^{(0)}$, although real, is \textit{not} symmetric, so that its nullspace \textit{differs} from its left-nullspace, with the latter being dependent on the follower load through the matrix $\bK_{\text{F}}(\bP)$.
When the follower loads vanish, $\bK_{\text{F}}(\bP)=\bzero$, matrix $\bA^{(0)}$ becomes symmetric, so that $\ker\trans{{\bA^{(0)}}}=\ker\bA^{(0)}$, and therefore its left-nullspace will contain only rigid-body translations.
This fact will be shown to have profound consequences both on the solvability of Eqs.~\eqref{eq:system_sequence} and on the symmetry of the resulting effective continuum material.

\subsection{The unsymmetric acoustic tensor for the hypo-elastic effective continuum}
\label{sec:acoustic_tensor_effective_continuum}
%
%
For Hermitian lattices the effective material behavior is uniquely defined by the equations of the asymptotic expansion~\eqref{eq:system_sequence} up to second-order~\cite{bordiga_2021}.
This result is now extended to non-Hermitian lattices loaded with follower forces.

Up to second-order, the matrices involved in Eqs.~\eqref{eq:system_sequence} are $\bA^{(0)}$, $\bA^{(1)}_{\bh}$, and $\bA^{(2)}_{\bh}$, which can be built through a series expansion of Eq.~\eqref{eq:system_matrix} as
\begin{equation}
    \label{eq:A0_A1_A2}
    \begin{aligned}
        \bA^{(0)}       & = \conjtrans{{\bZ^{(0)}}} \left( \bK^{(0)}_{\text{H}} + \bK_{\text{F}} \right) \bZ^{(0)} \,,                                                                                                      \\
        \bA^{(1)}_{\bh} & = \conjtrans{{\bZ^{(1)}_{\bh}}} \left( \bK^{(0)}_{\text{H}} + \bK_{\text{F}} \right) \bZ^{(0)} + \conjtrans{{\bZ^{(0)}}} \left( \bK^{(0)}_{\text{H}} + \bK_{\text{F}} \right) \bZ^{(1)}_{\bh} \,, \\
        \bA^{(2)}_{\bh} & = \conjtrans{{\bZ^{(2)}_{\bh}}} \left( \bK^{(0)}_{\text{H}} + \bK_{\text{F}} \right) \bZ^{(0)} + \conjtrans{{\bZ^{(0)}}} \left( \bK^{(0)}_{\text{H}} + \bK_{\text{F}} \right) \bZ^{(2)}_{\bh} +   \\
                        & \quad + \conjtrans{{\bZ^{(1)}_{\bh}}} \left( \bK^{(0)}_{\text{H}} + \bK_{\text{F}} \right) \bZ^{(1)}_{\bh} - \left( \omega^{(1)}_{\bh} \right)^2 \conjtrans{{\bZ^{(0)}}} \bM^{(0)} \bZ^{(0)} \,,
    \end{aligned}
\end{equation}
where $\bK^{(0)}_{\text{H}}$ and $\bM^{(0)}$ are the limits at vanishing frequency of $\bK_{\text{H}}$ and $\bM$, respectively, while $\bZ^{(j)}_{\bh}$ is the $j$-th term of the series expansion of $\bZ(\epsilon\,\bh)$ centered at $\epsilon=0$.
In the following, the notation $\bK^{(0)}=\bK^{(0)}_{\text{H}}+\bK_{\text{F}}$ is adopted for compactness.
By inspecting Eqs.~\eqref{eq:A0_A1_A2}, the following features emerge:
\begin{enumerate}[label=(\roman*)]
    \item $\bZ^{(j)}_{\bh}$ is real when $j$ is even and purely imaginary when $j$ is odd;
    \item The matrix of follower loads $\bK_{\text{F}}$ \textit{breaks} the \textit{symmetry} of $\bA^{(0)}$ and the \textit{anti-symmetry} of $\bA^{(1)}_{\bh}$;
    \item As for the Hermitian case, information about the eigenfrequencies starts to appear at second-order, where the square of the first-order term of the expansion~\eqref{eq:omega_expansion}, namely term $\left( \omega^{(1)}_{\bh} \right)^2$, appears as a factor multiplying the inertial term.
\end{enumerate}

Using expressions~\eqref{eq:A0_A1_A2}, the solution of Eqs.~\eqref{eq:system_sequence} can be constructed as follows.
At the lowest order, the only unknown is vector $\tilde{\bq}^{(0)}_{\bh}$, which satisfies the homogeneous linear system
\begin{equation}
    \label{eq:zeroth_order_system}
    \mO(\epsilon^0): \quad \conjtrans{{\bZ^{(0)}}} \bK^{(0)} \bZ^{(0)}\,\tilde{\bq}^{(0)}_{\bh} = \bzero \,.
\end{equation}
It is easy to recognize that vectors $\tilde{\bq}^{(0)}_{\bh}$ in the form of rigid-body translations are the only solutions of Eq.~\eqref{eq:zeroth_order_system}, so that\footnote{
    Note that Floquet-Bloch conditions imposed by the matrix $\bZ^{(0)}$ prevent rigid-body rotations from being admissible solutions of Eq.~\eqref{eq:zeroth_order_system}.
}
\begin{equation}
    \label{eq:q0}
    \tilde{\bq}^{(0)}_{\bh} = g_1\,\bt_1 + g_2\,\bt_2 \,,
\end{equation}
where the two rigid-body translations, parallel to vectors $\bt_1$ and $\bt_2$, and of magnitude $g_1$ and $g_2$ are assumed to be mutually orthogonal for convenience.
Coefficients $g_1$ and $g_2$ are to be determined through solution of the higher-order equations of system~\eqref{eq:system_sequence}.

The first-order expansion reported in Eqs.~\eqref{eq:system_sequence} is \textit{non-homogeneous}.
It becomes, with the help of Eq.~\eqref{eq:A0_A1_A2} and noticing that $\bK^{(0)} \bZ^{(0)}\,\tilde{\bq}^{(0)}_{\bh} = \bzero$ because $\tilde{\bq}^{(0)}_{\bh}$ is a rigid-body translation, the following expression
\begin{equation}
    \label{eq:first_order_system_simplified}
    \mO(\epsilon^1): \quad \conjtrans{{\bZ^{(0)}}} \bK^{(0)} \bZ^{(0)}\,\tilde{\bq}^{(1)}_{\bh} = -\conjtrans{{\bZ^{(0)}}} \bK^{(0)} \bZ^{(1)}_{\bh}\,\tilde{\bq}^{(0)}_{\bh} \,,
\end{equation}
whose only unknown is vector $\tilde{\bq}^{(1)}_{\bh}$, which is to be determined for every $\tilde{\bq}^{(0)}_{\bh}$ of representation~\eqref{eq:q0}.
As Eq.~\eqref{eq:first_order_system_simplified} is a non-homogeneous singular linear system, the solvability condition~\eqref{eq:solvability_condition} needs to be evaluated, which, specialized to Eq.~\eqref{eq:first_order_system_simplified}, assumes the form
\begin{equation}
    \label{eq:solvability_first_order_system}
    \by \scalp \conjtrans{{\bZ^{(0)}}} \bK^{(0)} \bZ^{(1)}_{\bh}\,\tilde{\bq}^{(0)}_{\bh} = 0 \,, \quad \forall \by\,\Big\lvert\,\conjtrans{{\bZ^{(0)}}} \trans{{\bK^{(0)}}} \bZ^{(0)} \by = \bzero \,.
\end{equation}
The importance of condition \eqref{eq:solvability_first_order_system} for non-Hermitian lattices can be appreciated by noting that this condition is trivially satisfied when follower loads are absent, i.e. for Hermitian lattices.
In fact, when $\bK_{\text{F}}=\bzero$, matrix $\bA^{(0)}$ is symmetric and thus the space of vectors $\by$ appearing in~ Eq.\eqref{eq:solvability_first_order_system} coincides with the space of rigid-body translations~\eqref{eq:q0}, so that condition~\eqref{eq:solvability_first_order_system} becomes
\begin{equation*}
    \bt_j \scalp \conjtrans{{\bZ^{(0)}}}\bK^{(0)}\bZ^{(1)}_{\bh}\,\tilde{\bq}^{(0)}_{\bh} =  \tilde{\bq}^{(0)}_{\bh} \scalp \trans{{\bZ^{(1)}_{\bh}}}\bK^{(0)}\bZ^{(0)}\,\bt_j = 0 \,, \quad \forall j\in\{1,2\} \,,
\end{equation*}
where the fact that $\bZ^{(0)}$ is real has been used.
On the contrary, for non-Hermitian lattices, the solvability of the first-order system~\eqref{eq:first_order_system_simplified} is not \textit{a priori} guaranteed, but has to be verified on a case-by-case basis, owing to the dependence of the left-nullspace of $\bA^{(0)}$ on the specific distribution of follower loads.

It is now demonstrated that, when the first-order system is solvable, the second-order system allows for the identification of the effective behavior in a way similar to the Hermitian case.
Let $\btau_1(\bh)$ and $\btau_2(\bh)$ be the two independent solutions of Eq.~\eqref{eq:first_order_system_simplified}, respectively corresponding to $\bt_1$ and $\bt_2$, so that any other solution $\tilde{\bq}^{(1)}_{\bh}$ can be expressed through the linear combination
\begin{equation}
    \label{eq:q1}
    \tilde{\bq}^{(1)}_{\bh} = g_1\,\btau_1(\bh) + g_2\,\btau_2(\bh) \,.
\end{equation}

Using Eqs.~\eqref{eq:A0_A1_A2}, the second-order system reported in Eqs.~\eqref{eq:system_sequence} becomes
\begin{equation}
    \label{eq:second_order_system}
    \begin{multlined}[0.9\linewidth]
        \mO(\epsilon^2): \quad \conjtrans{{\bZ^{(0)}}} \bK^{(0)} \bZ^{(0)}\,\tilde{\bq}^{(2)}_{\bh} =
        - \conjtrans{{\bZ^{(0)}}} \bK^{(0)} \bZ^{(2)}_{\bh}\,\tilde{\bq}^{(0)}_{\bh} + \\
        - \conjtrans{{\bZ^{(0)}}} \bK^{(0)} \bZ^{(1)}_{\bh}\,\tilde{\bq}^{(1)}_{\bh} - \conjtrans{{\bZ^{(1)}_{\bh}}} \bK^{(0)} \bZ^{(0)}\,\tilde{\bq}^{(1)}_{\bh} + \\
        - \conjtrans{{\bZ^{(1)}_{\bh}}} \bK^{(0)} \bZ^{(1)}_{\bh}\,\tilde{\bq}^{(0)}_{\bh} + \left( \omega^{(1)}_{\bh} \right)^2 \conjtrans{{\bZ^{(0)}}} \bM^{(0)} \bZ^{(0)}\,\tilde{\bq}^{(0)}_{\bh} \,,
    \end{multlined}
\end{equation}
where $\tilde{\bq}^{(0)}_{\bh}$ and $\tilde{\bq}^{(1)}_{\bh}$ are expressed as linear combinations defined by Eqs.~\eqref{eq:q0} and~\eqref{eq:q1}, respectively.
For simplicity, the following notation is introduced
\begin{equation}
    \label{eq:second_order_matrices}
    \begin{aligned}
        \bUpsilon & = \conjtrans{{\bZ^{(0)}}} \bK^{(0)} \bZ^{(1)}_{\bh} + \conjtrans{{\bZ^{(1)}_{\bh}}} \bK^{(0)} \bZ^{(0)} \,,       \\
        \bGamma   & = \conjtrans{{\bZ^{(0)}}} \bK^{(0)} \bZ^{(2)}_{\bh} + \conjtrans{{\bZ^{(1)}_{\bh}}} \bK^{(0)} \bZ^{(1)}_{\bh} \,, \\
        \bPsi     & = \conjtrans{{\bZ^{(0)}}} \bM^{(0)} \bZ^{(0)} \,,
    \end{aligned}
\end{equation}
so that, by letting $\by_1$ and $\by_2$ be the vectors spanning the left-nullspace of $\bA^{(0)}$ and using Eqs.~\eqref{eq:q0},~\eqref{eq:q1}, and~\eqref{eq:second_order_matrices}, the solvability conditions for the system~\eqref{eq:second_order_system} can be written as
\begin{equation}
    \label{eq:solvability_second_order_system}
    \begin{aligned}
        \by_1 \scalp \left( g_1 \left( \bGamma\,\bt_1 + \bUpsilon\,\btau_1 \right) + g_2 \left( \bGamma\,\bt_2 + \bUpsilon\,\btau_2 \right) - \left( \omega^{(1)}_{\bh} \right)^2 \bPsi\,(g_1\,\bt_1 + g_2\,\bt_2) \right) & = 0 \,, \\
        \by_2 \scalp \left( g_1 \left( \bGamma\,\bt_1 + \bUpsilon\,\btau_1 \right) + g_2 \left( \bGamma\,\bt_2 + \bUpsilon\,\btau_2 \right) - \left( \omega^{(1)}_{\bh} \right)^2 \bPsi\,(g_1\,\bt_1 + g_2\,\bt_2) \right) & = 0 \,.
    \end{aligned}
\end{equation}

Conditions~\eqref{eq:solvability_second_order_system} can be organized in a matrix form as
\begin{equation}
    \label{eq:system_effective_continuum}
    \left[\bXi - \left( \omega^{(1)}_{\bh} \right)^2\bPi \right] \bg = \bzero,
\end{equation}
where
\begin{equation*}
    \bXi=
    \begin{bmatrix}
        \by_1 \scalp \left( \bGamma\,\bt_1 + \bUpsilon\,\btau_1 \right) & \by_1 \scalp \left( \bGamma\,\bt_2 + \bUpsilon\,\btau_2 \right) \\
        \by_2 \scalp \left( \bGamma\,\bt_1 + \bUpsilon\,\btau_1 \right) & \by_2 \scalp \left( \bGamma\,\bt_2 + \bUpsilon\,\btau_2 \right)
    \end{bmatrix}, \quad
    \bPi=
    \begin{bmatrix}
        \by_1 \scalp \bPsi\,\bt_1 & \by_1 \scalp \bPsi\,\bt_2 \\
        \by_2 \scalp \bPsi\,\bt_1 & \by_2 \scalp \bPsi\,\bt_2
    \end{bmatrix},
\end{equation*}
which has the structure of an eigenvalue problem.
The eigenvalue $\left( \omega^{(1)}_{\bh} \right)^2$ and the corresponding eigenvector $\bg=\trans{\{g_1,g_2\}}$ represent the linearized dispersion relation and the polarization of the long-wavelength waveform, respectively (the lowest-order terms of the series expansion~\eqref{eq:omega_waveform_expansion}).
Thus Eq.~\eqref{eq:system_effective_continuum} defines the LF-LW behavior of the lattice structure, in other words, the behavior of the effective continuum.

System~\eqref{eq:system_effective_continuum} evidences the following features:
\begin{enumerate}[label=(\roman*)]
    \item Matrices $\bXi$ and $\bPi$ are real;
    \item Matrices $\bUpsilon$ and $\bGamma$ are, respectively, linear and quadratic forms in $\bh$, so that matrix $\bXi$ is in turn quadratic in $\bh$;
    \item Matrix $\bPi$ is independent of $\bh$, as a consequence of the independence of $\bPsi$ from $\bh$;
    \item Neither matrix $\bXi$ nor $\bPi$ is necessarily symmetric;
    \item\label{it:indeterminacy} Matrices $\bXi$ and $\bPi$ do \textit{not} result to be uniquely defined, because they depend upon the choice of the arbitrary vectors $\by_1$ and $\by_2$, which merely need to form a basis for the left-nullspace of $\bA^{(0)}$.
\end{enumerate}
The indeterminacy connected to the above point~\ref{it:indeterminacy} can be solved through a left-multiplication by $\rho\,\bPi^{-1}$ of the eigenvalue problem~\eqref{eq:system_effective_continuum}, which yields
\begin{equation}
    \label{eq:system_effective_continuum_normalized}
    \left( \hat{\bXi}\ - \rho \left( \omega^{(1)}_{\bh} \right)^2 \bI \right) \bg = \bzero \,,
\end{equation}
where $\hat{\bXi}=\rho\,\bPi^{-1}\bXi$, and $\rho$ is the average mass density of the lattice, $\rho = \frac{1}{\abs{\mC}} \sum_{j=1}^{N_b} \gamma_j l_j$, with $\abs{\mC}$ being the area of the unit cell.

The normalized problem~\eqref{eq:system_effective_continuum_normalized} has now exactly the same structure of the eigenvalue problem governing the wave propagation in a homogeneous elastic continuum, the so-called effective continuum.
This is because $\hat{\bXi}$ is a quadratic form in $\bh$, so that the eigenvalues $\omega^{(1)}_{\bh}$ are homogeneous functions of degree 1 with respect to $\bh$, characterizing the classical linear dispersion of an elastic solid.

The remarkable feature of the eigenvalue problem~\eqref{eq:system_effective_continuum_normalized} is the \textit{unsymmetry} of the \textit{effective acoustic tensor} $\hat{\bXi}$, which implies that
\begin{enumerate}[label=(\roman*)]
    \item \textit{the effective continuum is necessarily non-hyper-elastic}, as an hyper-elastic material must possess a symmetric acoustic tensor;
    \item \textit{complex conjugate eigenvalues cannot be excluded}, so that the effective solid can display a complex dispersion, which corresponds to flutter instability.
\end{enumerate}

The homogenization here presented demonstrates, in a constructive way, that non-hyper-elastic continua can actually arise from the LF-LW limit of lattice materials that incorporates non-conservative positional forces (follower loads).
It is also shown that the resulting effective material is still an incrementally \textit{linear} and \textit{elastic} material, meaning that the non-hyper-elasticity necessarily follows from the \textit{lack of major symmetry} of the effective constitutive operator, preventing the existence of an incremental strain-energy potential.

The following sections provide a concrete application of the presented analysis, investigate the stability properties of the non-hyper-elastic effective medium, and demonstrate that the latter behaves as an \textit{artificial material} capable of capturing a dynamic instability of the lattice that may be defined as \textit{macroscopic flutter}, corresponding to flutter at the scale of a continuum.

\section{Breaking selfadjointness in a lattice material}
\label{sec:grid_example}
\begin{figure}[htb]
    \centering
    \begin{subfigure}[b]{0.49\linewidth}
        \centering
        \includegraphics[width=0.98\linewidth]{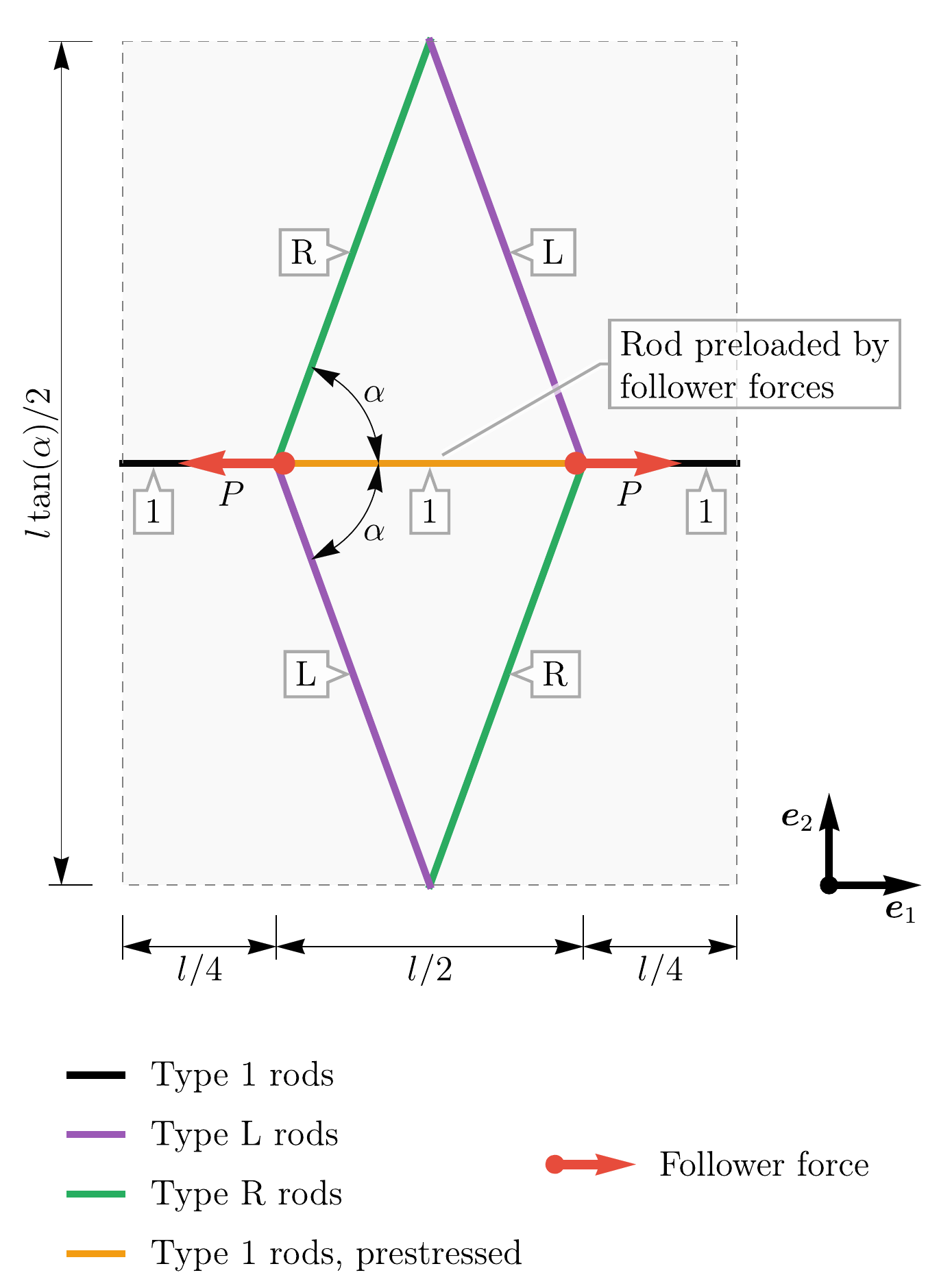}
    \end{subfigure}
    \begin{subfigure}[b]{0.49\linewidth}
        \centering
        \includegraphics[width=0.98\linewidth]{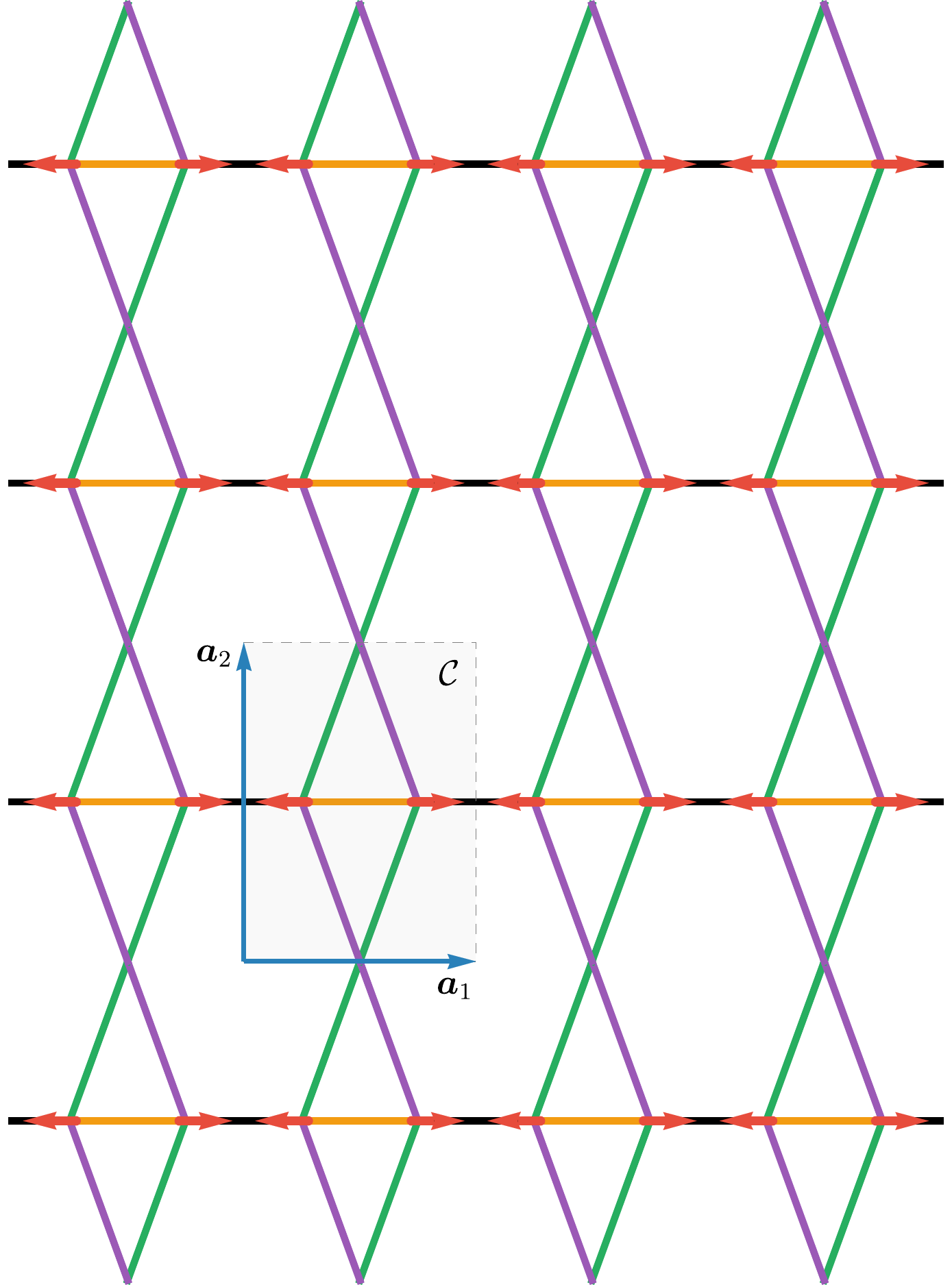}
    \end{subfigure}
    \caption{
        Anisotropic lattice preloaded by a periodic distribution of follower forces.
        The effective material following from homogenization is non-hyper-elastic and, at sufficiently high value of load, displays macroscopic flutter instabilities.
        The lattice is made up of three types of rods with different mechanical properties: 1 (black/orange), L (purple), and R (green).
    }
    \label{fig:geometry_lattice}
\end{figure}

%
%
The lattice configuration sketched in Fig.~\ref{fig:geometry_lattice} is used to demonstrate the existence of a hypo-elastic material response, even capable of displaying flutter instability.

The structure shown in the figure was optimized through an extensive search to identify the most suitable symmetry class that displays flutter, occurring as the \textit{critical} instability (meaning that static bifurcations do not occur before flutter).
Analysis of several structures indicates that flutter instability is critical for \textit{fully anisotropic} effective materials, while for orthotropic or cubic materials static instabilities (macroscopic or microscopic) usually occur before flutter.
This finding is in agreement with~\cite{bigoni_1999,piccolroaz_2006}, where, even though in an elastoplastic constitutive framework, anisotropy is shown to play an important role for the emergence of flutter instability.

The anisotropic lattice shown in Fig.~\ref{fig:geometry_lattice} is realized by assembling together rods of three types, labeled as 1, L, and R, which are allowed to have different values of mechanical parameters.
The geometry of the lattice can also be modified by means of the angle $\alpha\in(0^\circ, 90^\circ)$ of inclination between the rods of type 1 and type L or R.
The central rod 1 (colored in orange, Fig.~\ref{fig:geometry_lattice}) is assumed to be preloaded axially with two follower loads of magnitude $P$ applied at its ends.

In order to reduce the parameter space, the complete set of non-dimensional groups is identified, in addition to the angle $\alpha$, as follows.
\begin{subequations}
    \label{eq:nondimensional_groups}
    The dimensionless frequency and loading are defined as
    \begin{equation}
        \label{eq:nondimensional_omega_p}
        \Omega = \omega l \sqrt{\gamma_1/A_1} \,, \qquad p = \frac{P l^2}{2 B_1} \,,
    \end{equation}
    (where $\gamma_*$, $A_*$, and $B_*$ are respectively the linear mass density, the axial and the bending stiffness of the rod) while the mass density and stiffness of the rods form the following non-dimensional groups
    \begin{equation}
        \label{eq:nondimensional_mechanical_parameters}
        \begin{gathered}
            \mu_{\text{L}} = \gamma_{\text{L}}/\gamma_1 \,, \qquad \mu_{\text{R}} = \gamma_{\text{R}}/\gamma_1 \,, \qquad
            \chi_{\text{L}} = A_{\text{L}}/A_1 \,, \qquad \chi_{\text{R}} = A_{\text{R}}/A_1 \,, \\
            \lambda_1 = \frac{l}{2\sqrt{B_1/A_1}} \,, \qquad \lambda_{\text{L}} = \frac{l}{4\cos\alpha\sqrt{B_{\text{L}}/A_{\text{L}}}} \,, \qquad \lambda_{\text{R}} = \frac{l}{4\cos\alpha\sqrt{B_{\text{R}}/A_{\text{R}}}} \,,
        \end{gathered}
    \end{equation}
    where $\lambda_1$, $\lambda_{\text{L}}$, and $\lambda_{\text{R}}$ represent the slenderness of the three types of rods.
\end{subequations}

Results are presented for the following values
\begin{equation}
    \label{eq:nondimensional_mechanical_parameters_numeric}
    \mu_{\text{L}} = 1/25 \,, \,\,
    \mu_{\text{R}} = 3/10 \,, \,\,
    \chi_{\text{L}} = 1/25 \,, \,\,
    \chi_{\text{R}} = 1 \,, \,\,
    \lambda_1 = 40 \,, \,\,
    \lambda_{\text{L}} = 70 \,, \,\,
    \lambda_{\text{R}} = 40 \,, \,\,
    \alpha = 70^\circ ,
\end{equation}
while the loading parameter $p$ is left free to identify the condition for flutter instability.

The numerical values~\eqref{eq:nondimensional_mechanical_parameters_numeric} have been selected as representative of a structural grid that can, at least in principle, be realized in a laboratory experiment.

\subsection{Complex eigenvalues for the acoustic tensor of the effective medium}
\label{sec:complex_eigenvalues}
The asymptotic method outlined in Section~\ref{sec:homogenization} is directly applied to the anisotropic lattice reported in Fig.~\ref{fig:geometry_lattice}, so that the acoustic tensor $\hat{\bXi}$ governing the effective dynamic response of the material can be identified through Eq.~\eqref{eq:system_effective_continuum_normalized} as an explicit function of all the dimensionless parameters~\eqref{eq:nondimensional_groups}.
The numerical values~\eqref{eq:nondimensional_mechanical_parameters_numeric} are adopted, so that the eigenvalues of $\hat{\bXi}$ are reported as a function of only the propagation direction $\bn$ and the loading parameter $p$.
\begin{figure}[htb]
    \centering
    {\phantomsubcaption\label{fig:eigenvalues_acoustic_tensor_p0_re}}
    {\phantomsubcaption\label{fig:eigenvalues_acoustic_tensor_p10_re}}
    {\phantomsubcaption\label{fig:eigenvalues_acoustic_tensor_p60_re}}
    {\phantomsubcaption\label{fig:eigenvalues_acoustic_tensor_p0_im}}
    {\phantomsubcaption\label{fig:eigenvalues_acoustic_tensor_p10_im}}
    {\phantomsubcaption\label{fig:eigenvalues_acoustic_tensor_p60_im}}
    \includegraphics[width=\linewidth]{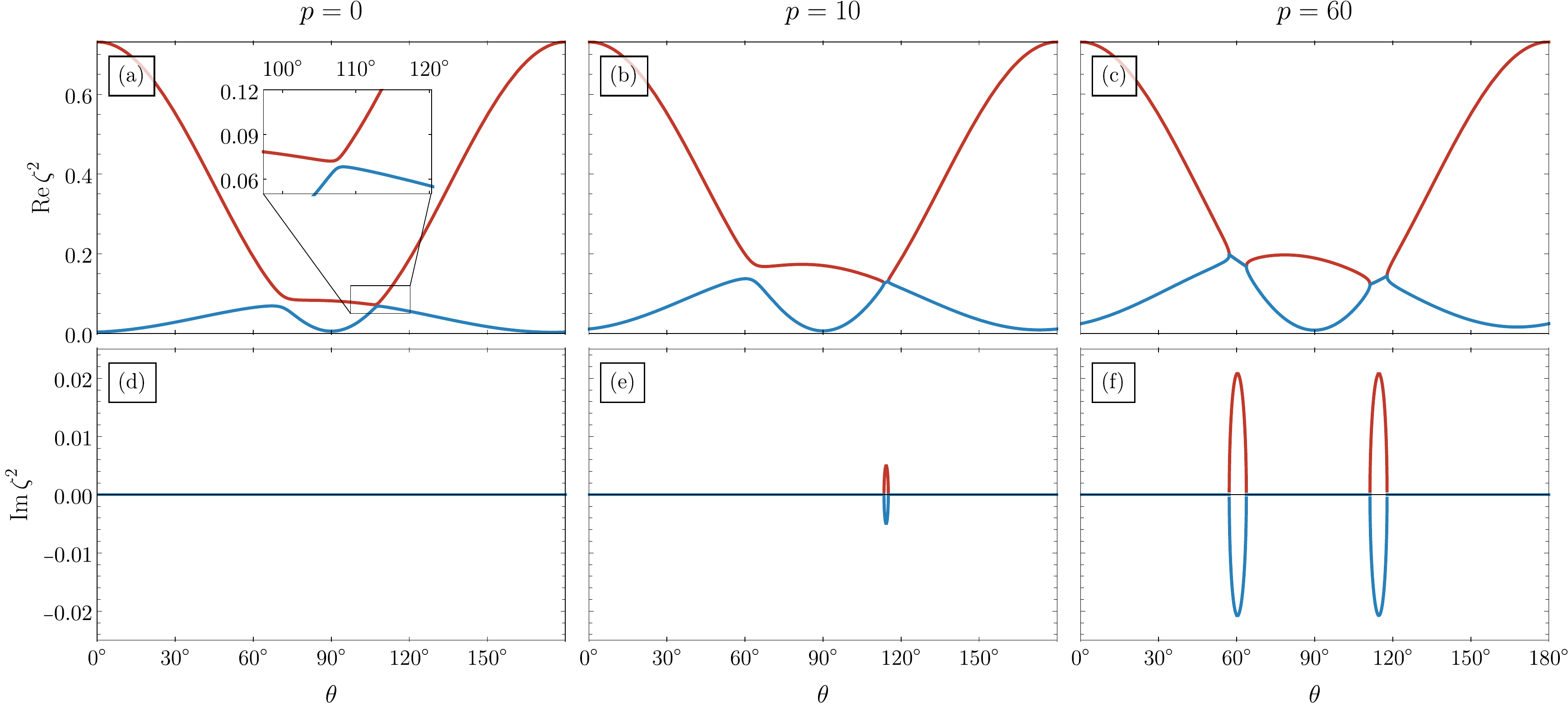}
    \caption{
        Real and imaginary parts of the dimensionless eigenvalues $\zeta^2$ of the acoustic tensor pertinent to an effective elastic material derived from the anisotropic lattice (Fig.~\ref{fig:geometry_lattice}).
        At null prestress, $p=0$, panels (\subref{fig:eigenvalues_acoustic_tensor_p0_re},~\subref{fig:eigenvalues_acoustic_tensor_p0_im}) the follower forces vanish, so that the symmetry of the acoustic tensor follows and the effective material is stable (i.e. $\zeta^2\in\Reals_+$).
        An increase of (tensile) axial load up to $p=10$ is sufficient to produce complex conjugate eigenvalues in a narrow range of directions centered around $\theta \approx 115^\circ$, panels (\subref{fig:eigenvalues_acoustic_tensor_p10_re},~\subref{fig:eigenvalues_acoustic_tensor_p10_im}).
        This range is broadened when the load is increased up to $p=60$, panels (\subref{fig:eigenvalues_acoustic_tensor_p60_re},~\subref{fig:eigenvalues_acoustic_tensor_p60_im}), where a second range of complex conjugate eigenvalues appears centered around $\theta \approx 60^\circ$.
    }
    \label{fig:eigenvalues_acoustic_tensor}
\end{figure}

Eq.~\eqref{eq:system_effective_continuum_normalized} is conveniently made dimensionless as follows
\begin{equation}
    \label{eq:system_effective_continuum_normalized_dimensionless}
    \frac{l}{A_1 \norm{\bh}^2}\left( \hat{\bXi}(\bh)\ - \rho \left( \omega^{(1)}_{\bh} \right)^2 \bI \right) \bg = \bzero \,,
\end{equation}
so that the propagation direction can be identified as $\bn=\bh/\norm{\bh}=\cos\theta\,\be_1+\sin\theta\,\be_2$ and the dimensionless eigenvalues can be defined by
\begin{equation}
    \label{eq:dimensionless_eigenvalues}
    \zeta^2 = \frac{\rho\, l}{A_1 \norm{\bh}^2} \left( \omega^{(1)}_{\bh} \right)^2 \,,
\end{equation}
where the average mass density for the lattice under study is
\begin{equation*}
    \rho = \frac{\mu_{\text{L}} + \mu_{\text{R}} + 2 \cos\alpha}{\sin\alpha}\frac{\gamma_1}{l} \approx 1.09\,\frac{\gamma_1}{l} \,.
\end{equation*}

The nature of the eigenvalues $\zeta^2$ of the acoustic tensor dictates the stability of the effective medium which can be analyzed as a function of the loading parameter $p$.
Fig.~\ref{fig:eigenvalues_acoustic_tensor} reports results for the solutions $\zeta^2(\theta)$ at three different values of loading.
At vanishing axial load, $p=0$, (Fig.~\ref{fig:eigenvalues_acoustic_tensor_p0_re},~\subref{fig:eigenvalues_acoustic_tensor_p0_im}) the follower forces are absent, so that the acoustic tensor is symmetric and the effective material is stable (i.e. $\zeta^2\in\Reals_+$).
An increase in the magnitude of the (tensile) follower loads up to $p=10$ is sufficient to generate complex conjugate eigenvalues in a narrow range of directions centered around $\theta \approx 115^\circ$, see Fig.~\ref{fig:eigenvalues_acoustic_tensor_p10_re},~\subref{fig:eigenvalues_acoustic_tensor_p10_im}.
This range is broadened by further increasing the load up to $p=60$ (Fig.~\ref{fig:eigenvalues_acoustic_tensor_p60_re},~\subref{fig:eigenvalues_acoustic_tensor_p60_im}), a level which also generates a second range of complex conjugate eigenvalues centered around $\theta \approx 60^\circ$.

The results shown in Fig.~\ref{fig:eigenvalues_acoustic_tensor} can be synthesized in the following statements, never previously pointed out:
\begin{enumerate}[label=(\roman*)]
    \item The lattice material behaves as a non-hyper-elastic effective medium;
    \item Flutter instability of the effective solid can be triggered by a \textit{tensile and follower} preload.
\end{enumerate}
It follows that a new and completely unexplored way is disclosed towards the realization of \textit{elastic} materials without a strain-energy potential.
As shown in Section~\ref{sec:energy_loop}, these materials are able to exchange energy with the environment by undergoing closed strain cycles of non-vanishing area.

\subsection{Flutter domains vs ellipticity loss in the effective medium}
\label{sec:flutter_domains}
%
%
The identification of the critical flutter load for the lattice follows from a computation of domains in the space $\{\theta,\, p\}$, where the eigenvalues~\eqref{eq:dimensionless_eigenvalues} becomes
complex conjugate, $\Ip\zeta^2 \neq 0$.
This computation provides, in addition to the flutter load, also the fan of directions along which the blowing-up waves characterizing flutter do appear in the forced response of the material (see Section~\ref{sec:flutter_forced}).

\begin{figure}[htb]
    \centering
    {\phantomsubcaption\label{fig:flutter_domain_80deg}}
    {\phantomsubcaption\label{fig:flutter_domain_70deg}}
    {\phantomsubcaption\label{fig:flutter_domain_60deg}}
    {\phantomsubcaption\label{fig:flutter_domain_50deg}}
    \includegraphics[width=0.9\linewidth]{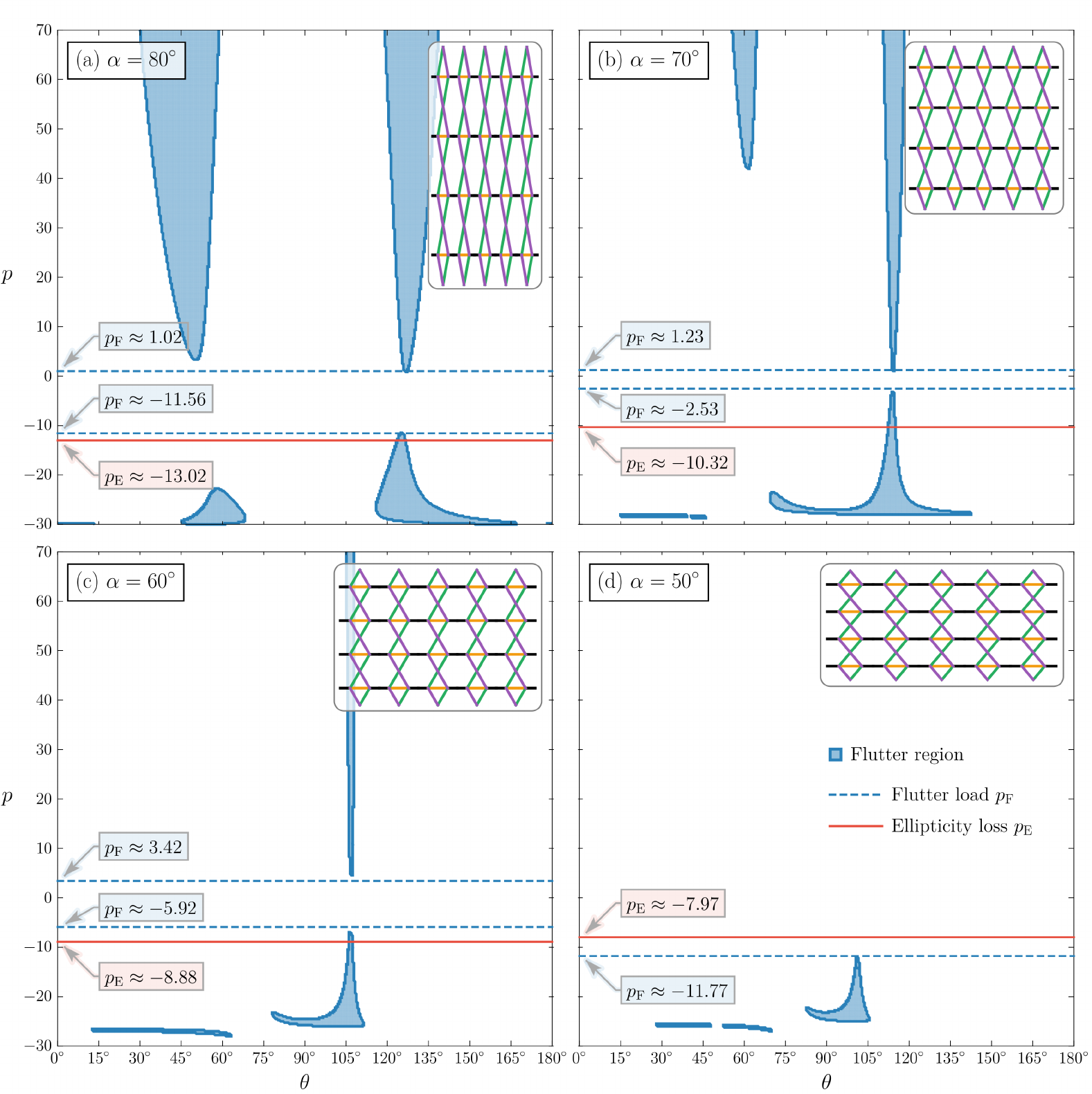}
    \caption{
        Flutter domains (blue regions in the load $p$ vs wave inclination $\theta$ plane) and threshold for ellipticity loss (red lines) are reported for an elastic material `equivalent' to the anisotropic lattice of Fig.~\ref{fig:geometry_lattice} at different values of angle $\alpha$.
        Note the strong dependence of the flutter regions on the inclination angle $\alpha$, showing that the tensile flutter regions ($p>0$) become narrower as $\alpha$ is decreased from $80^\circ$ to $60^\circ$, until tensile flutter load does not occur at $\alpha=50^\circ$, panel~(\subref{fig:flutter_domain_50deg}).
        A flutter load in compression ($p<0$) is present in all four panels, with panels~(\subref{fig:flutter_domain_80deg}),~(\subref{fig:flutter_domain_70deg}), and~(\subref{fig:flutter_domain_60deg}) showing that \textit{flutter occurs before ellipticity loss} (as for compression $\abs{p_{\text{F}}}<\abs{p_{\text{E}}}$), while for $\alpha=50^\circ$ panel~(\subref{fig:flutter_domain_50deg}) shows that \textit{flutter cannot be reached within the elliptic regime} (as $\abs{p_{\text{F}}}>\abs{p_{\text{E}}}$).
    }
    \label{fig:flutter_domains}
\end{figure}

Fig.~\ref{fig:flutter_domains} shows flutter domains in the load $p$ vs wave inclination $\theta$ plane, denoted by blue regions and computed for the parameter set~\eqref{eq:nondimensional_mechanical_parameters_numeric} at different values for the angle $\alpha\in\{50^\circ,\, 60^\circ,\, 70^\circ,\, 80^\circ\}$.
In the same figure, the threshold identifying loss of ellipticity (denoted with continuous red lines) is also reported.
Ellipticity loss corresponds to the condition of localization of deformation, a quasi-static material instability.
Parts~\ref{fig:flutter_domain_80deg},~\subref{fig:flutter_domain_70deg}, and~\subref{fig:flutter_domain_60deg} demonstrate that the effective medium can reach flutter instability for both compressive and tensile loads.
However, for compression, $p<0$, the threshold for loss of ellipticity becomes close to the flutter load, so that only a small part of the flutter region is attainable in compression before a macroscopic static bifurcation occurs (see for example Fig.~\ref{fig:flutter_domain_80deg} and~\subref{fig:flutter_domain_60deg}).

\textit{
    The effective material remains, instead, elliptic under a tensile follower loading ($p>0$), so that the flutter regions can be reached without prior encountering any static bifurcations.
}

Moreover, the dependence of the flutter regions on the angle $\alpha$ is quite strong, as demonstrated by the severe narrowing of the flutter region in tension caused by the decrease of $\alpha$ from $80^\circ$ (Fig.~\ref{fig:flutter_domain_80deg}) to $70^\circ$ (Fig.~\ref{fig:flutter_domain_70deg}), as well as by the complete disappearance of tensile flutter loads for $\alpha=50^\circ$.
Fig.~\ref{fig:flutter_domain_50deg} also shows that flutter is still present in compression, but the region \textit{cannot} be attained without first encountering a loss of ellipticity.

These results demonstrates the extreme tunability of the anisotropic lattice, whose geometry can be designed to reach a critical flutter instability, both in tension and compression, as well as to avoid flutter and to meet the condition of ellipticity loss, corresponding to the localization of deformation.

\subsection{Complex-valued lattice dispersion vs the effective medium}
\label{sec:lattice_dispersion}
%
%
So far the effective medium has been the main subject of the investigation, providing evidence that a macroscopic flutter instability can be attained for the anisotropic lattice under study.
In this section, a proof is provided that the prediction of the effective material are actually accurate for capturing the LF-LW behavior of the lattice.
This is conducted by comparing directly the dispersion relation of the lattice (simulated numerically via finite elements, using COMSOL Multiphysics\textsuperscript{\textregistered}) to the wave speeds predicted by the effective medium.
Furthermore, to better characterize the dynamic behavior of the lattice also at medium/high frequency, the actual nonlinear lattice dispersion is analyzed through the computation of dispersion diagrams and dispersion surfaces.

\begin{figure}[htb]
    \centering
    {\phantomsubcaption\label{fig:dispersion_diagrams_p0_re}}
    {\phantomsubcaption\label{fig:dispersion_diagrams_p10_re}}
    {\phantomsubcaption\label{fig:dispersion_diagrams_p60_re}}
    {\phantomsubcaption\label{fig:dispersion_diagrams_p0_im}}
    {\phantomsubcaption\label{fig:dispersion_diagrams_p10_im}}
    {\phantomsubcaption\label{fig:dispersion_diagrams_p60_im}}
    \includegraphics[width=\linewidth]{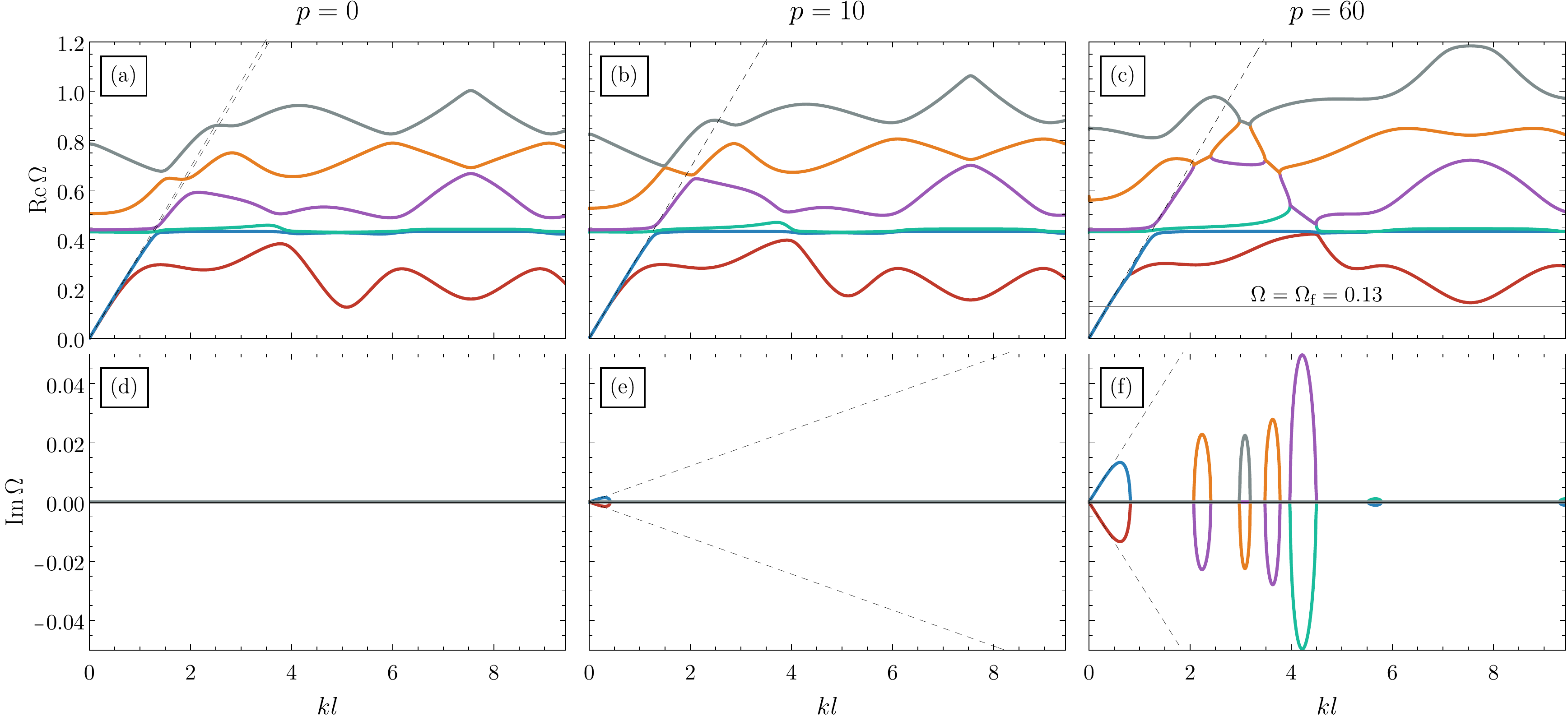}
    \caption{
        Complex dispersion diagrams (only the first six branches are reported, ordered by the real part), computed along the direction of the wave vector $\bk = k \left( \cos\theta\,\be_1 + \sin\theta\,\be_2 \right)$ with $\theta=2\,\text{rad}\approx 115^\circ$, for the anisotropic lattice (Fig.~\ref{fig:geometry_lattice}).
        Colored solid lines denote the dispersion relation for the lattice (numerically simulated via finite elements), while the dashed black lines refer to the effective medium.
        The two diagrams perfectly match within LF-LW regime.
        The gray horizontal line in~(\subref{fig:dispersion_diagrams_p60_re}) indicates the frequency $\Omega_{\text{f}}=0.13$ chosen for the analysis of the forced response reported in Fig.~\ref{fig:forced_response_flutter}.
        As the follower loading is increased from $p=0$, panels (\subref{fig:dispersion_diagrams_p0_re},~\subref{fig:dispersion_diagrams_p0_im}), complex conjugate branches appear in the LF-LW region at $p=10$, panels (\subref{fig:dispersion_diagrams_p10_re},~\subref{fig:dispersion_diagrams_p10_im}), as well as at higher frequency when $p=60$, panels (\subref{fig:dispersion_diagrams_p60_re},~\subref{fig:dispersion_diagrams_p60_im}).
    }
    \label{fig:dispersion_diagrams}
\end{figure}

Fig.~\ref{fig:dispersion_diagrams} shows the first six branches of the dispersion diagram pertaining to the lattice, computed along a fixed direction in the wave vector space defined by $\bk = k \left( \cos\theta\,\be_1 + \sin\theta\,\be_2 \right)$ with $\theta=2\,\text{rad}\approx 115^\circ$.
This direction corresponds to flutter for the lattice inclination $\alpha=70^\circ$ at a critical load $p_{\text{F}}=1.23$ (Fig.~\ref{fig:flutter_domain_70deg}) and is selected to examine the `nucleation' of the complex-valued dispersion that the effective medium predicts to occur when the follower load is increased from zero.
The dispersion diagrams is evaluated in the dimensionless space $\{kl, \Omega\}$.

The figure shows the great accuracy of the homogenization, confirmed by the perfect match between the LF-LW dispersion curves (solid colored lines) and the linear dispersion of the non-hyper-elastic effective solid (dashed black lines).
Furthermore, as the loading parameter increases from $p=0$ (Fig.~\ref{fig:dispersion_diagrams_p0_re},~\subref{fig:dispersion_diagrams_p0_im}) to $p=10$ (Fig.~\ref{fig:dispersion_diagrams_p10_re},~\subref{fig:dispersion_diagrams_p10_im}), the dispersion relation becomes complex-valued in the LF-LW region (first and second branch near $k l = 0$), but remains real-valued at higher frequency.
A further increase of load up to $p=60$ (Fig.~\ref{fig:dispersion_diagrams_p60_re},~\subref{fig:dispersion_diagrams_p60_im})
\begin{enumerate*}[label=(\roman*)]
    \item leads to a higher slope of the imaginary part of both the first and the second branch emanating from the origin, and
    \item causes the higher branches also to become complex-valued within some intervals of the normalized wave number $kl$.
\end{enumerate*}
\begin{figure}[htbp]
    \centering
    {\phantomsubcaption\label{fig:dispersion_surfaces_p0_re}}
    {\phantomsubcaption\label{fig:dispersion_surfaces_p0_im}}
    {\phantomsubcaption\label{fig:dispersion_surfaces_p10_re}}
    {\phantomsubcaption\label{fig:dispersion_surfaces_p10_im}}
    {\phantomsubcaption\label{fig:dispersion_surfaces_p60_re}}
    {\phantomsubcaption\label{fig:dispersion_surfaces_p60_im}}
    \includegraphics[width=0.85\linewidth]{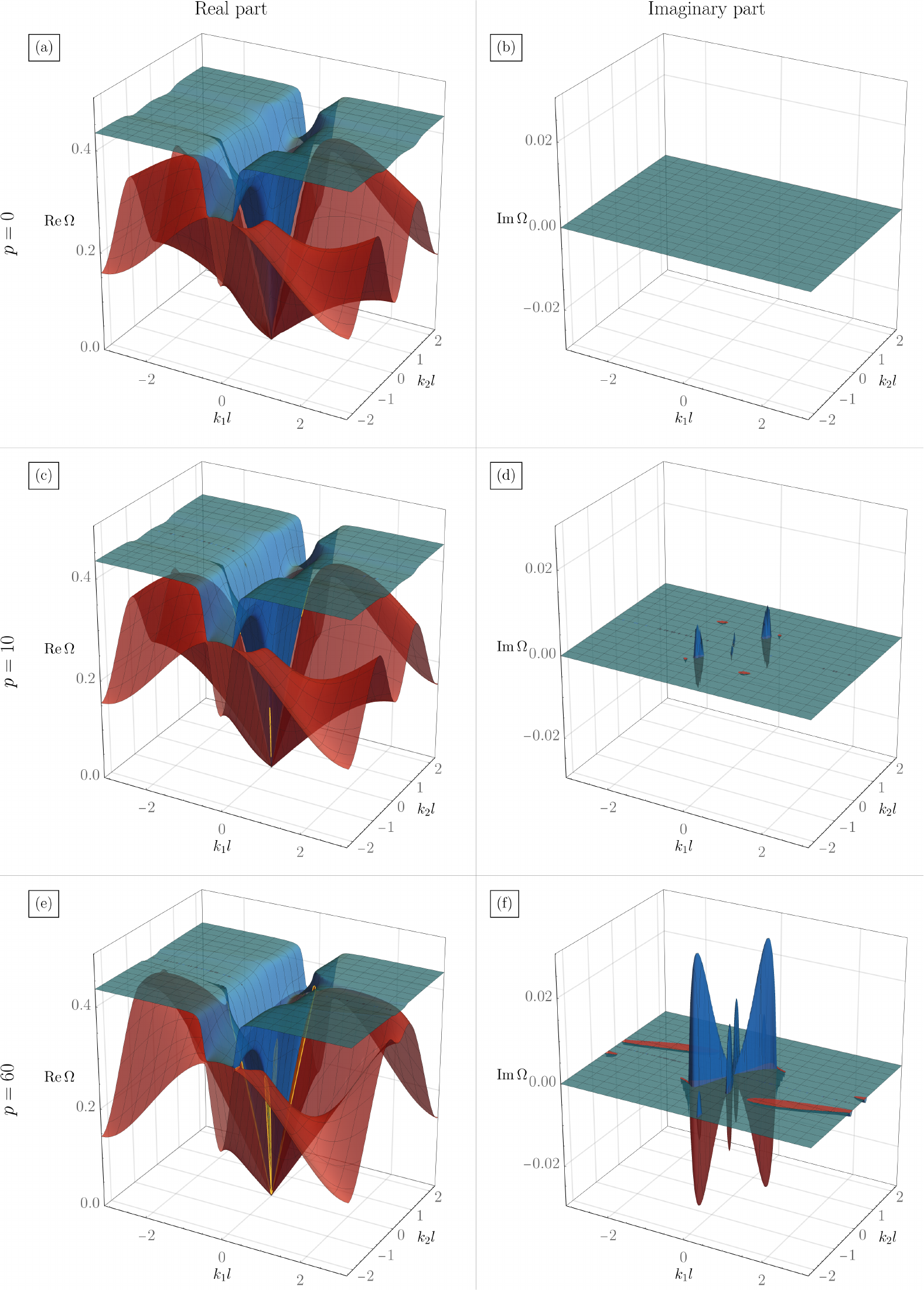}
    \caption{
        Complex dispersion surfaces of the acoustic branches for the anisotropic lattice (Fig.~\ref{fig:geometry_lattice}), subject to three different levels of follower loading $p=0, 10, 60$.
        Parts~(\subref{fig:dispersion_surfaces_p0_re}) and~(\subref{fig:dispersion_surfaces_p0_im}) refer to the case of a stable Hermitian lattice so that the dispersion surfaces have vanishing imaginary part.
        As $p$ is increased, parts~(\subref{fig:dispersion_surfaces_p10_re})--(\subref{fig:dispersion_surfaces_p60_im}), the acoustic branches become complex-valued in a complex-conjugate fashion.
        In parts~(\subref{fig:dispersion_surfaces_p10_re})--(\subref{fig:dispersion_surfaces_p60_re}), the zones where the surfaces of the real part `merge' are highlighted in yellow.
    }
    \label{fig:dispersion_surfaces}
\end{figure}
\begin{figure}[htbp]
    \centering
    {\phantomsubcaption\label{fig:dispersion_surfaces_p60_re_1}}
    {\phantomsubcaption\label{fig:dispersion_surfaces_p60_re_2}}
    {\phantomsubcaption\label{fig:dispersion_slowness_p60_re_1}}
    {\phantomsubcaption\label{fig:dispersion_slowness_p60_re_2}}
    {\phantomsubcaption\label{fig:dispersion_surfaces_p60_im_1}}
    {\phantomsubcaption\label{fig:dispersion_surfaces_p60_im_2}}
    {\phantomsubcaption\label{fig:dispersion_slowness_p60_im_1}}
    {\phantomsubcaption\label{fig:dispersion_slowness_p60_im_2}}
    \includegraphics[width=0.85\linewidth]{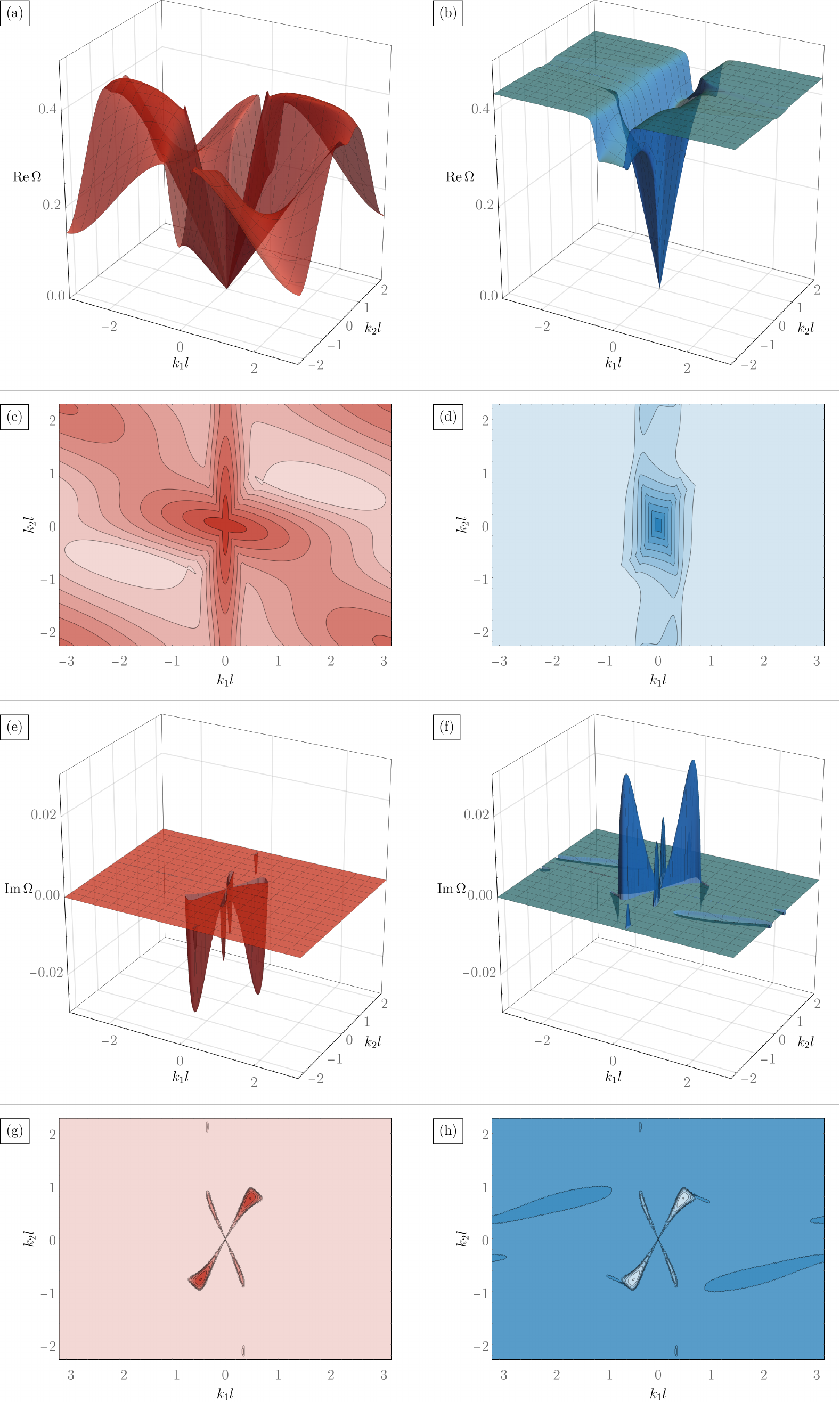}
    \caption{
        Real and imaginary parts of the complex dispersion surfaces and complex slowness contours for the two acoustic branches of the anisotropic lattice, subject to a follower loading of intensity $p=60$.
        The severe and localized gradients that can be observed in the real parts, panels (\subref{fig:dispersion_surfaces_p60_re_1})--(\subref{fig:dispersion_slowness_p60_re_2}), are the typical signature of a strong dynamic anisotropy of the lattice.
        Parts~(\subref{fig:dispersion_surfaces_p60_im_1})--(\subref{fig:dispersion_slowness_p60_im_2}) show the structure of the imaginary part of the acoustic branches emanating from the origin of the $\bk$-space that are responsible for the macroscopic flutter instability.
    }
    \label{fig:dispersion_surfaces_slowness_p60}
\end{figure}

The dispersion diagrams reported in Fig.~\ref{fig:dispersion_diagrams} are useful for visualizing the formation of the complex-valued dispersion, but they reveal the structure of the dispersion branches only along a predefined path in the $\bk$-space.
In order to fully appreciate the two-dimensional features of the dispersion relation, the complex-valued eigenfrequencies $\Omega(\bk)$ corresponding to the two acoustic branches have been computed on the full reciprocal unit cell, for $-\pi<k_1 l<\pi$ and $-2\pi\cot\alpha<k_2 l<2\pi\cot\alpha$, and the real and imaginary parts are plotted in Fig.~\ref{fig:dispersion_surfaces} as 3D plots in the $\Omega$--$\bk l$ dimensionless space to visualize the \textit{complex dispersion surfaces}.

The imaginary part of the dispersion surfaces (Fig.~\ref{fig:dispersion_surfaces}) clearly shows the formation of the complex-valued acoustic branches, triggered by the increase of the follower load from $p=0$, up to $p=10$ and beyond, $p=60$.
Furthermore, the `merging' of the real parts can be appreciated from yellow parts (looking like lines, but having a finite area) depicted in the 3D views of Fig.~\ref{fig:dispersion_surfaces_p60_re}, to indicate the presence of complex-conjugate eigenfrequencies.
In order to provide a more distinct visualization of the complex structure of the acoustic branches, a detailed representation of the real and imaginary parts is reported in Fig.~\ref{fig:dispersion_surfaces_slowness_p60} for $p=60$, where the dispersion surfaces are plotted separately along with the corresponding slowness contours.

Fig.~\ref{fig:dispersion_surfaces_p60_re_1}--\subref{fig:dispersion_slowness_p60_re_2} shows that the real part of the dispersion surfaces is characterized by severe and localized gradients, representing the typical signature of a strong dynamic anisotropy of the system.
On the other hand, Fig.~\ref{fig:dispersion_surfaces_p60_im_1}--\subref{fig:dispersion_slowness_p60_im_2} shows the presence of `islands' of wave vectors for which there is a non-vanishing imaginary part of the eigenfrequencies, thus revealing the directions and wavelengths of the blowing-up wave modes, characteristic of flutter instability.

It is also worth noting that the dispersion surfaces provide a valuable information about the presence of static bifurcation of both macro and microscopic type~\cite{bordiga_2021}.
As can be seen in Fig.~\ref{fig:dispersion_surfaces}, the applied tensile follower loading does \textit{not} cause any eigenfrequency to vanish for any non-vanishing wave vector, indicating that short wavelength static bifurcations are excluded.
Moreover, the slope of the acoustic branches is non-vanishing for every propagation direction, so that the effective material remains within the elliptic range and macroscopic static bifurcations are also excluded.

The presented dispersion analysis, besides demonstrating the accuracy of the non-hyper-elastic effective medium in the LF-LW regime, suggests also a way to trigger the macroscopic flutter instability through the application of an external pulsating load (the dynamic Green's function approach~\cite{piccolroaz_2006}).
It is in fact expected that, when the forcing frequency is selected within the range of complex-valued acoustic branches (for some directions of propagation), the response of the lattice will be characterized by waves displaying an exponential amplification, as shown in~\cite{piccolroaz_2006}.
Furthermore, the case $p=60$ (Fig.~\ref{fig:dispersion_diagrams_p60_re},~\subref{fig:dispersion_diagrams_p60_im}) shows the simultaneous presence of multiple complex-valued branches, in addition to those present in the LF-LW region.
It may therefore be speculated that a `high-frequency flutter' might interact with macroscopic flutter, occurring at low frequency, when a dynamic excitation is applied to the lattice.
This interaction does not in fact occur, as will be shown in Section~\ref{sec:flutter_forced} through numerical investigation of the forced response of the lattice.

It is also worth recalling that the complex-valued dispersion found here is not due to the presence of viscosity or thermo-mechanical coupling, which simply entail a temporal damping of the material response (see for instance~\cite{hussein_2009,frazier_2016,bacigalupo_2019a}) and thus cannot produce any flutter instability.

\section{Disclosing flutter at the macroscopic scale}
\label{sec:flutter_forced}
%
%
The manifestation of macroscopic flutter is disclosed by analyzing the time-harmonic response of the effective medium corresponding, via homogenization, to the infinite anisotropic lattice shown in Fig.~\ref{fig:geometry_lattice}, when a concentrated force is applied, pulsating at sufficiently low frequency, as proposed in~\cite{piccolroaz_2006}.

The lattice response is also computed numerically through time-harmonic finite element simulations, used to compare with the response of the infinite-body Green's function.
F.E. calculations have been performed in COMSOL Multiphysics\textsuperscript{\textregistered} on a finite-size square domain $350\,l$ wide, thus having 350 unit cells along the $\be_1$ direction.
\begin{figure}[htb]
    \centering
    {\phantomsubcaption\label{fig:forced_response_no_flutter_G1_lattice}}
    {\phantomsubcaption\label{fig:forced_response_no_flutter_G2_lattice}}
    {\phantomsubcaption\label{fig:forced_response_no_flutter_G1_continuum}}
    {\phantomsubcaption\label{fig:forced_response_no_flutter_G2_continuum}}
    \includegraphics[width=0.9\linewidth]{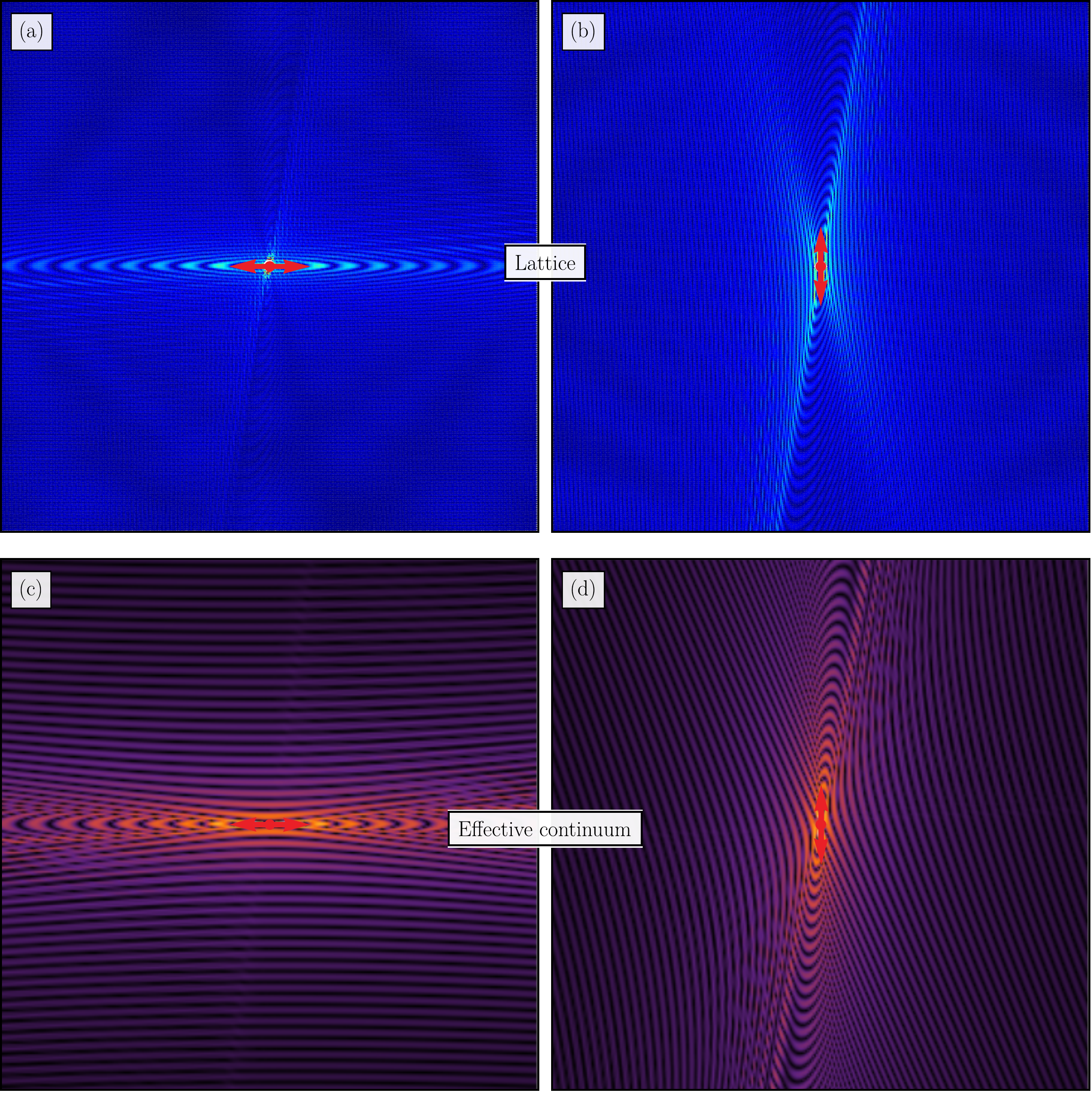}
    \caption{
        Time-harmonic forced response of the anisotropic lattice (upper part) and of the corresponding effective continuum (lower part), displaying a stable behavior for $p = 1 < p_{\text{F}}$ (colors based on the norm of real part of the displacement field).
        The material does not evidence flutter instability, so that the response is strongly anisotropic, but \textit{decays} away from the pulsating load.
    }
    \label{fig:forced_response_no_flutter}
\end{figure}
The square domain is surrounded by an additional layer, $50\,l$ wide, endowed with viscous damping which is properly tuned to minimize reflections back to the inner region (thus acting as an \textit{ad-hoc} PML).
Each rod has been discretized using standard Euler-Bernoulli elements (cubic shape functions) with the maximum dimension $l/20$.
The computational mesh has been selected after testing three different element sizes, $l/30$, $l/20$, and $l/10$, where $l/20$ has been found to be the appropriate trade-off between accuracy and computational cost.
The incremental contribution of the follower loads is introduced by adding the terms given by Eq.~\eqref{eq:total_incremental_work_follower} directly to the weak form of the equations of motion of the lattice.

Two cases of follower load are considered, $p=1$ and $p=60$, respectively below and above the critical load for macroscopic flutter $p_{\text{F}} \approx 1.23$ (Fig.~\ref{fig:flutter_domain_70deg}).
The forcing frequency is set equal to $\Omega=\Omega_{\text{f}}=0.13$, while the other lattice parameters are defined by the set of values~\eqref{eq:nondimensional_mechanical_parameters_numeric}.
The response of the lattice is compared to the time-harmonic Green's function response of a hypo-elastic material, endowed with the effective non-symmetric acoustic tensor found via homogenization.
The time-harmonic Green's function for an arbitrary non-symmetric constitutive law has been derived in~\cite{piccolroaz_2006} and is not repeated here.
Introducing the components $n_1$ and $n_2$ of the unit wave propagation vector $\bn$, the effective acoustic tensor computed through the homogenization technique results as follows (reported with 4 significant figures):
\begin{itemize}
    \item For $p=1<p_{\text{F}}$
          \begin{equation*}
              \begin{aligned}
                  \hat{\bXi}(\bn) = & \left[\left( 0.7310 n_1^2 + 0.004622 n_1 n_2 + 0.006831 n_2^2 \right) \be_1\otimes\be_1 +    \right.   \\
                                    & + \left( 0.002312 n_1^2 + 0.02823 n_1 n_2 + 0.01370 n_2^2 \right) \be_1\otimes\be_2 +                  \\
                                    & + \left( -0.003075 n_1^2 + 0.02825 n_1 n_2 + 0.01345 n_2^2 \right) \be_2\otimes\be_1 +                 \\
                                    & \left.+ \left( 0.007138 n_1^2 + 0.02734 n_1 n_2 + 0.1615 n_2^2 \right) \be_2\otimes\be_2 \right] A_1/l \\
              \end{aligned}
          \end{equation*}
    \item For $p=60>p_{\text{F}}$
          \begin{equation*}
              \begin{aligned}
                  \hat{\bXi}(\bn) = & \left[\left( 0.7318 n_1^2 + 0.009171 n_1 n_2 + 0.01370 n_2^2 \right) \be_1\otimes\be_1 +    \right.       \\
                                    & + \left( 0.004606 n_1^2 + 0.04323 n_1 n_2 + 0.03802 n_2^2 \right) \be_1\otimes\be_2 +                     \\
                                    & + \left( -0.1946 n_1^2 + 0.03928 n_1 n_2 + 0.02572 n_2^2 \right) \be_2\otimes\be_1 +                      \\
                                    & \left.+ \left( 0.02300 n_1^2 + 0.06475 n_1 n_2 + 0.1832 n_2^2 \right) \be_2\otimes\be_2 \right] A_1/l \,. \\
              \end{aligned}
          \end{equation*}
\end{itemize}
Comparisons between lattice and continuum are performed by excluding a small disk of radius $4\,l$, cut around the loading point to avoid the `blinding' effect caused by the logarithmic singularity inherent to the Green's function.
\begin{figure}[htb]
    \centering
    {\phantomsubcaption\label{fig:forced_response_flutter_G1_lattice}}
    {\phantomsubcaption\label{fig:forced_response_flutter_G2_lattice}}
    {\phantomsubcaption\label{fig:forced_response_flutter_G1_continuum}}
    {\phantomsubcaption\label{fig:forced_response_flutter_G2_continuum}}
    \includegraphics[width=0.9\linewidth]{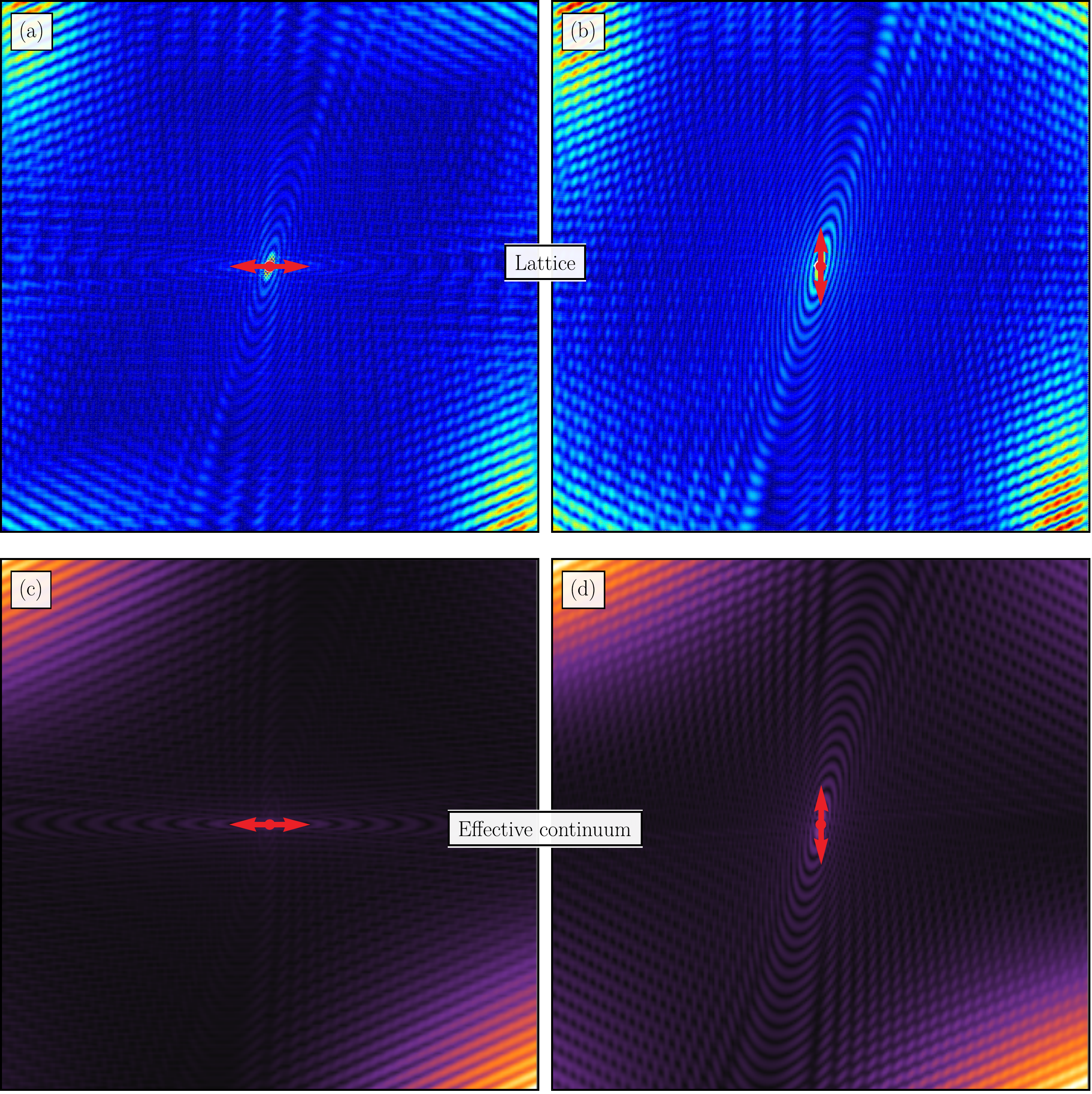}
    \caption{
        Time-harmonic forced response of the anisotropic lattice (upper part) and of the corresponding effective continuum (lower part), displaying flutter instability at $p = 60 > p_{\text{F}}$ (colors based on the norm of real part of the displacement field).
        The response is typical of the behavior of a dynamically unstable material, evidencing blowing-up waves traveling away from the forcing source, but localized along two preferential directions. The latter are well captured by the spectral analysis of the effective acoustic tensor, characterized by complex-conjugate eigenvalues (Figs.~\ref{fig:eigenvalues_acoustic_tensor_p60_re},~\subref{fig:eigenvalues_acoustic_tensor_p60_im} and~\ref{fig:flutter_domain_70deg}).
    }
    \label{fig:forced_response_flutter}
\end{figure}

Results pertaining to the case $p = 1 < p_{\text{F}}$ are reported in Fig.~\ref{fig:forced_response_no_flutter}, showing colored density plots of the magnitude of the normalized displacement field (only the real part is reported).
In this first figure, the follower loading is not high enough to induce complex-conjugate eigenvalues for the effective acoustic tensor.
Therefore, flutter instability does not occur and the response is strongly anisotropic (compare parts~\subref{fig:forced_response_no_flutter_G1_lattice},~\subref{fig:forced_response_no_flutter_G1_continuum} to~\subref{fig:forced_response_no_flutter_G2_lattice},~\subref{fig:forced_response_no_flutter_G2_continuum}), but \textit{decays} away from the pulsating load, revealing the signature of a dynamically stable material.

The decaying response reported in Fig.~\ref{fig:forced_response_no_flutter} is in stark contrast to the results shown in Fig.~\ref{fig:forced_response_flutter}, pertaining to the case $p = 60 > p_{\text{F}}$.
Here the follower loading is above the critical value for flutter instability and therefore the forced response reveals the presence of wave characterized by amplitudes exponentially growing away from the forcing source.
The directions of these blowing-up waves is in agreement with the fan of angles for which the effective acoustic tensor displays complex-conjugate eigenvalues (Fig.~\ref{fig:eigenvalues_acoustic_tensor_p60_re},~\subref{fig:eigenvalues_acoustic_tensor_p60_im} and Fig.~\ref{fig:flutter_domain_70deg}).
In particular, the magnitude of the displacement field decays in the immediate neighborhood of the loading point, but exponentially grows at a sufficient distance and along the predicted fan of directions.
Moreover, the effect related to the direction of application of loading (horizontal in Fig.~\ref{fig:forced_response_flutter_G1_lattice},~\subref{fig:forced_response_flutter_G1_continuum} and vertical in Fig.~\ref{fig:forced_response_no_flutter_G2_lattice},~\subref{fig:forced_response_no_flutter_G2_continuum}) is evident in a region close to the loading point, while at a further distance, information about the perturbing agent is `obscured' by the rapidly growing response.
This suggests that \textit{the key factor for triggering flutter instability is the state of the material and not the nature of the disturbance, a feature typical of a material instability}.
\begin{figure}[htbp]
    \centering
    {\phantomsubcaption\label{fig:fourier_tranform_G_1_p_1}}
    {\phantomsubcaption\label{fig:fourier_tranform_G_2_p_1}}
    {\phantomsubcaption\label{fig:fourier_tranform_G_1_p_60}}
    {\phantomsubcaption\label{fig:fourier_tranform_G_2_p_60}}
    \includegraphics[width=0.9\linewidth]{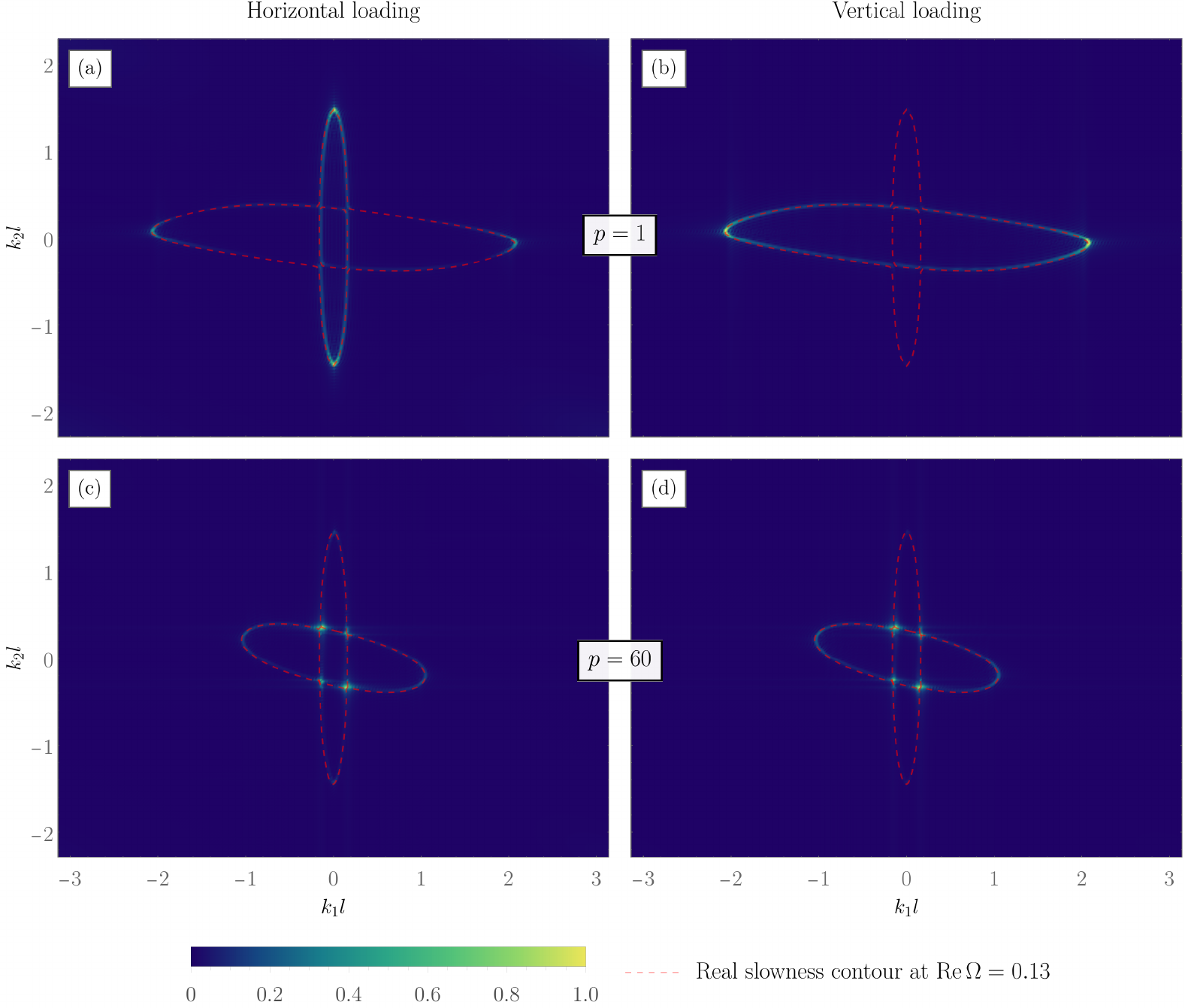}
    \caption{
        Fourier transform of the lattice forced response with the superimposed real slowness contour corresponding to the forcing angular frequency $\Rp\Omega=\Omega_f=0.13$.
        Parts~(\subref{fig:fourier_tranform_G_1_p_1},~\subref{fig:fourier_tranform_G_2_p_1}) correspond to the stable response of Fig.~\ref{fig:forced_response_no_flutter_G1_lattice},~\subref{fig:forced_response_no_flutter_G2_lattice}, while parts~(\subref{fig:fourier_tranform_G_1_p_60},~\subref{fig:fourier_tranform_G_2_p_60}) refer to the fluttering response of Fig.~\ref{fig:forced_response_flutter_G1_lattice},~\subref{fig:forced_response_flutter_G2_lattice}.
        The amplitude of the Fourier tranform shows that, for the stable case~(\subref{fig:fourier_tranform_G_1_p_1},~\subref{fig:fourier_tranform_G_2_p_1}), the response is composed of waves extracted from a quite diffused part of the slowness contour, while an increase of the loading up to $p=60$~(\subref{fig:fourier_tranform_G_1_p_60},~\subref{fig:fourier_tranform_G_2_p_60}) causes a sharp concentration of the excited waves along the parts of the slowness contour where the acoustic branches `merge' to become complex-conjugate.
    }
    \label{fig:fourier_tranform}
\end{figure}

Fig.~\ref{fig:forced_response_flutter} demonstrates an excellent agreement between the lattice response and the Green's function of the effective medium, thus proving that the effective behavior of a lattice material loaded by follower forces can be adequately described by a hypo-elastic, not hyper-elastic, continuum.
This, in turn, also suggests that the interaction between macroscopic flutter and `high-frequency flutter' does not occur.
In fact, Fig.~\ref{fig:forced_response_flutter} shows that, when the forcing frequency is sufficiently low, the possible complex-valued higher branches of the dispersion relation (e.g. those shown in Fig.~\ref{fig:dispersion_diagrams_p60_re},~\subref{fig:dispersion_diagrams_p60_im}) do not alter the manifestation of the flutter instability at the scale of the continuum.

An additional proof of the fact that the lattice is undergoing a macroscopic type of flutter instability can be obtained by probing the wave spectrum of the lattice response via the computation of the corresponding Fourier transform.
By denoting the transform of the horizontal and vertical displacement fields as $\mF(u_1)$ and $\mF(u_2)$ (efficiently evaluated through FFT), Fig.~\ref{fig:fourier_tranform} shows the density plot of the quantity $\sqrt{\abs{\mF(u_1)}^2+\abs{\mF(u_2)}^2}$, providing a measure of the displacement amplitude in the Fourier space.

Fig.~\ref{fig:fourier_tranform} depicts a perfect superposition between the set of dominant waves (marked in green/yellow) and the real slowness contour, computed at the forcing angular frequency $\Rp\Omega=\Omega_f=0.13$, thus verifying the accuracy of the numerical simulations.
By comparing results referred to the stable case (Fig.~\ref{fig:fourier_tranform_G_1_p_1},~\subref{fig:fourier_tranform_G_2_p_1}) and those pertaining to the flutter instability (Fig.~\ref{fig:fourier_tranform_G_1_p_60},~\subref{fig:fourier_tranform_G_2_p_60}), it can be observed that, as the loading is increased from $p=1$ to $p=60$, the spectrum of excited waves concentrates onto the parts of slowness contour where the acoustic branches `merge' to become complex-conjugate.
Hence, the Fourier tranform confirms that Bloch modes characterized by complex-valued dispersion dominate the response of the lattice undergoing flutter instability.

\section{Conclusions}
\label{sec:conclusions}
Is it possible to design an artificial material, which is neither subject to actuation, nor temperature changes or electromagnetic fields, and which is merely hypo-elastic and does not admit an elastic potential?
A positive answer to this question has been provided in this article.
The material, obtained as the effective response of a lattice of elastic rods prestressed with follower forces, is passive, but at the same time may absorb energy from the environment, by performing closed cycles in the strain space, a feature crucial for energy harvesting applications.
The obtained hypo-elastic material has been shown to be subject to flutter instability, consisting in an unbounded growth of propagating waves.
Remarkably, flutter is found not only for compressive, but also for \textit{tensile} follower loads.
On the one hand, this instability has been so far considered impossible for elastic materials, while on the other it has been advocated, but never experimentally found or theoretically proven, for plastic materials characterized by the non-associative flow rule.
Therefore, this article proves that flutter instability is possible in an elastic continuum, thus closing a problem remained open since at least 50 years and opening a new way to the design of unexplored materials.

\section*{Funding}
\label{sec:funding}
G.B. and A.P. gratefully acknowledges the funding from the European Union’s Horizon 2020 research and innovation programme under the Marie Sklodowska-Curie grant agreement No 955944-REFRACTURE2. D.B. acknowledges the funding from the European Union’s Horizon 2020 research and innovation programme under the Marie Sklodowska-Curie grant agreement No PITN-GA-2019-813424-INSPIRE.

\appendix

\section{Short review on non-Hermitian systems}
\label{sec:non-hermitian_literature}
Non-Hermitian response, generally associated to lack of reciprocity, has been in recent years the focus of an intense research interest, embracing optical~\cite{xu_2016,el-ganainy_2018,zhao_2019}, quantum~\cite{bender_1998,gong_2018}, acoustical~\cite{fleury_2015}, and mechanical~\cite{brandenbourger_2019,ghatak_2020,rosa_2020} systems. In statics, non-reciprocity violates the Betti identity, while in dynamics, non-reciprocity breaks the invariance occurring when source and receiver are swapped in a medium.
To eliminate this invariance in elastodynamics or to violate the Betti identity in statics, various systems have been introduced, which are listed below.
\begin{itemize}
    \item `Kinetic media' contain moving parts or circulating flows, a simple example being a fluid in constant motion, see for instance~\cite{attarzadeh_2018}.
          In these media reciprocity is lost as a consequence of a sort of Doppler effect, which violates the source/receiver invariance, but has nothing to do with the concept pointed out in the present article.
    \item Nonlinear materials do not obey the Betti reciprocity theorem, which holds in a linear theory.
          Regarding nonlinear materials, however, a possible correspondence with the results presented in this article would be a lack of the principal symmetry of the linearized material response, governed by an elastic fourth-order operator.
          This operator has never been explored so far as connected to the lack of reciprocity in dynamics.
          The so-called `static non-reciprocity'~\cite{wallen_2018} is known in mechanics since a long time as `bimodularity' of the response of a material, see among many others~\cite{shapiro_1966,green_1977,benveniste_1980}.
          Bimodularity can be caused by unilateral contact, or by the attainment of a bifurcation or an instability load.
          Bimodularity also occurs in a wire (in a masonry), which can sustain tension (compression), but not compression (tension).
          Bimodularity has nothing to do with the lack of the major symmetry of the elastic fourth-order operator and is therefore not pertinent to the results presented in this article.
    \item Activated materials change their constitutive properties in time (or in space) in response to an external stimulus~\cite{torrent_2018,torrent_2018a,wang_2018,chen_2019,attarzadeh_2020}.
          Usually these systems are 1D chains, for instance~\cite{brandenbourger_2019,ghatak_2020}, and therefore do not correspond to elastic \textit{two-dimensional materials} (even if it can be speculated that the 1D concept could be suitably generalized to more complex set-ups).
          Non-reciprocity effects can also be found in media designed to create topologically protected wave propagation, for instance introducing gyroscopes in a lattice~\cite{garau_2018,attarzadeh_2019,carta_2020,nieves_2020}, or non-local feedback controls~\cite{rosa_2020}.
          In the former case Ziegler~\cite{ziegler_1977} has shown that gyroscopic forces cannot destabilize a conservative linear system, so that a material equivalent to the discrete system analyzed in those papers would be crucially different from that found in the present article, which is shown to display flutter instability.
          The latter case where feedback controls are introduced in discrete systems is leading to the results closer to those reported here.
          However, in both cases homogenization of the response into an effective elastic medium has not been attempted.
\end{itemize}

It is remarked that the non-Hermitian response addressed in the present article has a precise technical meaning, namely, that the effective fourth-order elastic tensor lacks the so-called `major symmetry' (while the minor symmetries are never met in the incremental theory considered here, based on the increment of the first Piola-Kirchhoff stress and the gradient of incremental deformation).
In this sense, all the above-mentioned papers are, strictly speaking, not pertinent because there is no comparison with an effective elastic medium.
In addition, our system is \textit{passive, not active}.
The generation of the follower forces used in the present article has been shown to be possible with passive systems involving Coulomb friction~\cite{bigoni_2018a} or non-holonomic constraints~\cite{cazzolli_2020}.

\printbibliography

\end{document}